\documentclass{article}

\pdfoutput=1 

    \PassOptionsToPackage{numbers, sort&compress}{natbib}

\usepackage{filecontents}  

\usepackage[final]{neurips_2020}

\usepackage[utf8]{inputenc} 
\usepackage[T1]{fontenc}    
\usepackage{hyperref}       
\usepackage{url}            
\usepackage{booktabs}       
\usepackage{amsfonts}       
\usepackage{nicefrac}       
\usepackage{microtype}      

\usepackage{graphicx}
\graphicspath{{./graphics/}}
\usepackage{algorithm}
\usepackage[noend]{algorithmic}
\usepackage{subfigure}
\usepackage{multirow}
\usepackage{amsmath}
\usepackage{amssymb}
\usepackage[standard]{ntheorem}
\usepackage[inline]{enumitem}
\usepackage[inline=true,margin=false]{fixme}
\usepackage{xcolor}
\usepackage{makecell}
\usepackage{wrapfig}

\usepackage{xr}
\newtheorem{thm}{Theorem}

\theorembodyfont{\upshape}

\renewcommand{\vec}[1]{\boldsymbol{\mathbf{#1}}}

\title{CHIP: A Hawkes Process Model for Continuous-time Networks with Scalable and Consistent Estimation}

\author{%
  Makan Arastuie \\
  EECS Department\\
  University of Toledo\\
  \texttt{makan.arastuie}\\
  \texttt{@rockets.utoledo.edu}\\
  \And
  Subhadeep Paul\\
  Department of Statistics\\
  The Ohio State University\\
  \texttt{paul.963@osu.edu}\\
  \And
  Kevin S. Xu\\
  EECS Department\\
  University of Toledo\\
  \texttt{kevin.xu@utoledo.edu}\\
}

\begin{document}

\maketitle

\begin{abstract}
In many application settings involving networks, 
such as messages between users of an on-line social network or transactions between traders in financial markets, 
the observed data consist of timestamped relational events, which form a continuous-time network.
We propose the \emph{Community Hawkes Independent Pairs (CHIP)} generative model for such networks. 
We show that applying spectral clustering to an aggregated adjacency matrix constructed from the CHIP model provides \emph{consistent community detection} for a growing number of nodes and time duration. 
We also develop consistent and computationally efficient estimators for the model parameters.
We demonstrate that our proposed CHIP model and estimation procedure scales to large networks with tens of thousands of nodes and provides superior fits than existing continuous-time network models on several real networks. 
\end{abstract}

\section{Introduction}
A variety of complex systems in the computer, information, biological, and social sciences can be represented as a network, which consists of a set of objects (nodes) and relationships (edges) between the objects. 
In many application settings, we observe 
edges in the form of distinct events occurring between nodes over time.
For example, in on-line social networks, users interact with each other through events that occur at specific time instances such as liking, mentioning, or sharing another user's content. 
Such interactions form \emph{timestamped relational events}, where each event is a triplet $(i,j,t)$ denoting events from node $i$ (sender) to node $j$ (receiver) at timestamp $t$. The observation of these triplets defines a dynamic network that continuously evolves over time.

Timestamped relational event data are usually modeled by combining a point process model for the event times with a network model for the sender and receiver \citep{DuBois2010,Blundell2012,Dubois2013,xin2015continuous,matias2015semiparametric,yang2017decoupling, miscouridou2018modelling,corneli2018multiple,junuthula2017block}. 
We refer to such models as \emph{continuous-time network models} because they provide probabilities of observing events between two nodes during arbitrarily short time intervals. 
For model-based exploratory analysis and prediction of future events with relational event data, continuous-time network models are often superior to their discrete-time counterparts
\citep{xing10,Yang2011,Xu2014a,Xu2015,Matias2016}, 
which first aggregate events over time windows to form discrete-time network ``snapshots'' and thus lose granularity in modeling temporal dynamics.

We propose the \emph{Community Hawkes Independent Pairs (CHIP)} model, which is inspired by the recently proposed Block Hawkes Model (BHM)  \citep{junuthula2017block}  for timestamped relational event data. Both CHIP and BHM are based on the Stochastic Block Model (SBM) for static networks \citep{holland1983stochastic}. 
In the BHM, events between different pairs of nodes belonging to the same pair of communities are dependent, which makes it difficult to analyze. In contrast, for CHIP the pairs of nodes in the same community generate events according to \emph{independent} univariate Hawkes processes with shared parameters, so that the number of parameters remains the same as in the BHM. 
The independence between node pairs enables tractable analysis of the CHIP model and more scalable estimation than the BHM.

Our main contributions are as follows. 
(1) We demonstrate that spectral clustering provides consistent community detection in the CHIP model for a growing number of nodes and time duration. 
(2) We propose consistent and computationally efficient estimators for the model parameters also for a growing number of nodes and time duration.
(3) We show that the CHIP model provides better fits to several real datasets and scales to much larger networks than existing models, including a Facebook network with over $40,\!000$ nodes and over $800,\!000$ events.
Other point process network models have demonstrated good empirical results, but to the best of our knowledge, this work provides the \emph{first  theoretical guarantee of estimation accuracy}.
Our asymptotic analysis also has tremendous practical value given the scalability of our model to large networks with tens of thousands of nodes.

\section{Background}

\subsection{Hawkes Processes} \label{sec:HawkesProcesses}

The Hawkes process \cite{hawkes1971spectra} is a counting process designed to model continuous-time arrivals of events that naturally cluster together in time, where the arrival of an event increases the chance of the next event arrival immediately after. They have been used to model  earthquakes \citep{marsan2008extending}, financial markets \citep{embrechts2011multivariate,bacry2015hawkes}, and user interactions on social media 
\citep{Dubois2013, zhou2013learning}.

A univariate Hawkes process is a \emph{self-exciting} point process where its conditional intensity function given a sequence of event arrival times $\{t_1, t_2, t_3, ..., t_l\}$ for $l$ events up to time duration $T$ takes the general form
$\lambda(t) = \mu + \sum_{t_i<t}^{t_l}{\gamma(t - t_i)},$ 
where $\mu$ is the background intensity and $\gamma(\cdot)$ is the kernel or the excitation function. 
A frequent choice of kernel is an exponential kernel, parameterized by $\alpha, \beta > 0$ as $\gamma(t - t_i) = \alpha e^{-\beta (t - t_i)}$, where the arrival of an event instantaneously increases the conditional intensity by the jump size $\alpha$, after which the intensity decays exponentially back towards $\mu$ at rate $\beta$. Restricting $\alpha < \beta$ yields a stationary process.
We use an exponential kernel for the CHIP model, since it has been shown to provide a good fit for relational events in social media \citep{halpin2013modelling, masuda2013self, zhao2015seismic, junuthula2017block}.

\subsection{The Stochastic Block Model}
\label{sec:sbm}

Statistical models for networks typically 
consider a static network rather than a network of relational events. 
Many static network models are discussed in the survey by 
\citet{goldenberg2010survey}. 
A static network with $n$ nodes can be represented by an $n \times n$ adjacency matrix $A$ where $A_{ij}=1$ if there is an edge between nodes $i$ and $j$ and $A_{ij}=0$ otherwise. 
We consider networks with no self-edges, so $A_{ii}=0$ for all $i$. 
For a directed network, we let $A_{ij}=1$ if there is an edge from node $i$ to node $j$.

One model that has received significant attention 
is the \emph{stochastic block model} (SBM), formalized by \citet{holland1983stochastic}.
In the SBM, every node $i$ is assigned to one and only one community or \emph{block} $c_i \in \{1, \ldots, k\}$, where $k$ denotes the total number of blocks.
For a directed SBM, given the block membership vector $\vec{c} = [c_i]_{i=1}^n$, all off-diagonal entries of the adjacency matrix $A_{ij}$ are independent Bernoulli random variables with parameter $p_{c_i,c_j}$, where $p$ is a $k\times k$ matrix of probabilities. 
Thus the probability of forming an edge between nodes $i$ and $j$ depends only on the block memberships $c_i$ and $c_j$.
There have been significant advancements in the analysis of estimators for the SBM.
Several variants of spectral clustering \citep{von2007tutorial}, including regularized versions \cite{chaudhuri2012spectral, amini2013pseudo}, have been shown to be consistent estimators of the community assignments in the SBM and various extensions in several asymptotic settings \citep{rcy11,Sussman2012,qin13,lei2015consistency,chin2015stochastic,han2015consistent,joseph2016impact,gao2017achieving,vu2018simple,zhang2018understanding,bhattacharyya2018spectral,pensky2019spectral}. 
Spectral clustering scales to large networks with tens of thousands of nodes and is generally not sensitive to initialization, so it is also a practically useful estimator.

\subsection{Related Work}
One approach for modeling continuous-time networks is to treat the edge strength of each node pair as a continuous-time function that increases when an event occurs between the node pair and then decays afterwards \citep{jin2001structure, ahmad2018tie, zuo2019models}. 
Another approach is to combine a point process model for the event times, typically some type of Hawkes process, with a network model. 
The conditional intensity functions of the point processes then serve as the time-varying edge strengths. 
Point process network models are used in two main settings. 
The first involves estimating the structure of a latent or unobserved network from observed events at the nodes 
\citep{Linderman2014,linderman2015scalable,Farajtabar2015,Tran2015,He2015,hall2014tracking}. 
These models are often used to estimate \emph{static} networks of diffusion from information cascades.

In the second setting, which we consider in this paper, we directly observe events \emph{between pairs of nodes} so that events take on the form $(i,j,t)$ denoting an event from node $i$ to node $j$ at timestamp $t$. 
Our objective is to model the dynamics of such event sequences. 
In many applications, including messages on on-line social networks, most pairs of nodes either never interact and thus have no events between them. 
Thus, most prior work in this setting utilizes low-dimensional latent variable representations of the networks to parameterize the point processes. 

The latent variable representations are often inspired by generative models for static networks such as continuous latent space models \citep{hrh02} and stochastic block models \citep{holland1983stochastic}, resulting in the development of point process network models with continuous latent space representations \citep{yang2017decoupling} and latent block or community representations \citep{DuBois2010,Blundell2012,Dubois2013,xin2015continuous,matias2015semiparametric,junuthula2017block, miscouridou2018modelling,corneli2018multiple}, respectively. 
Point process network models with latent community representations are most closely related to the model we consider in this paper.
Exact inference in such models is intractable due to the discrete nature of the community assignments. 
Approximate inference techniques including Markov Chain Monte Carlo (MCMC) \citep{Blundell2012,Dubois2013,miscouridou2018modelling} or variational inference \citep{matias2015semiparametric,junuthula2017block,corneli2018multiple} have been used in prior work. 
While such techniques have demonstrated good empirical results, to the best of our knowledge, they come with no theoretical guarantees.

\section{The Community Hawkes Independent Pairs (CHIP) Model}

We consider a generative model for timestamped relational event networks that we call the \emph{Community Hawkes Independent Pairs (CHIP)} model. 
The CHIP model has parameters ($\vec{\pi}, \mu,\alpha, \beta$). 
Each node is assigned to a community or block $a \in \{1, \ldots, k\}$ with probability $\pi_a$, where each entry of $\vec{\pi}$ is non-negative and all entries sum to $1$. 
We represent the block assignments of all nodes either by a length $n$ vector $\vec{c} = [c_i]_{i=1}^n$ or an $n\times k$ binary matrix $C$ where $c_i = q$ is equivalent to $C_{iq} = 1$, $C_{iq'} = 0$ for all $q' \neq q$. 
Each of the parameters $\mu, \alpha, \beta$ is a $k \times k$ matrix. 
While we assume that the number of blocks and the block assignments of the nodes do not change with time, the CHIP model captures time-varying behavior due to the incorporation of self-exciting point processes.
Event times between node pairs $(i,j)$ within a block pair $(a,b)$ follow independent exponential Hawkes processes with shared parameters: baseline rate $\mu_{ab}$, jump size $\alpha_{ab}$, and decay rate $\beta_{ab}$. 
The generative process for our proposed CHIP model is as follows:
\begin{align*}  
c_i &\sim \text{Categorical}(\vec{\pi}) & & \text{for all nodes } i\\
\vec{t}_{ij} &\sim \text{Hawkes process}(\mu_{c_i c_j},\alpha_{c_i c_j},\beta_{c_i c_j}) & & \text{for all } i \neq j \\
Y &= \text{Row concatenate }[(i\vec{1},j\vec{1},\vec{t}_{ij})] & & \text{over all } i \neq j
\end{align*}
$\vec{1}$ denotes the all-ones vector of appropriate length. 
Let $T$ denote the end time of the Hawkes process, which would correspond to the duration of the data trace. 
The column vector of event times $\vec{t}_{ij}$ has length $N_{ij}(T)$, which denotes the number of events from node $i$ to node $j$ up to time $T$. Let $Y$ denote the event matrix with dimension $l \times 3$, where $l = \sum_{i,j} N_{ij}(T)$ denotes the total number of observed events over all node pairs. 
It is constructed by row concatenating triplets $(i,j,t_{ij}(q))$ over all events $q \in \{1, \ldots, N_{ij}(T)\}$ for all node pairs $i, j \in \{1, \ldots, n\}, i \neq j$.

\subsection{Relation to Other Models}
\label{sec:relation}
Our proposed CHIP model has a generative structure inspired by the SBM for static networks. 
Other point process network models in the literature have also utilized similar block structures,
but they have been incorporated in two different approaches.
One approach involves placing point process models at the level of block pairs \citep{Blundell2012,xin2015continuous,matias2015semiparametric,junuthula2017block}. 
For a network with $k$ blocks, $k^2$ different point processes are used to generate events between the $k^2$ block pairs. 
To generate events between pairs of nodes, rather than pairs of blocks, the point processes are thinned by randomly selecting nodes from the respective blocks so that all nodes in a block are stochastically equivalent, in the spirit of the SBM. 
Such models have demonstrated good empirical results, but the dependency between node pairs complicates analysis of the models. 

The other approach involves modeling pairs of nodes with independent point processes that share parameters among nodes in the same block \citep{DuBois2010,Dubois2013, corneli2018multiple}. 
By having node pairs in the same block share parameters, the number of parameters is the same as for the models with block pair-level point processes. 
However, by using independent point processes for all node pairs, there is no dependency between node pairs, which simplifies analysis of the model. 
We use this approach in the proposed CHIP model and exploit this independence to perform the theoretical analysis in Section \ref{sec:analysis}.

\subsection{Estimation Procedure} \label{sec:EstimationProcedure}

\begin{algorithm}[tb]
  \caption{Estimation procedure for Community Hawkes Independent Pairs (CHIP) model}
  \label{alg:estimation}
  
  \begin{algorithmic}
    \STATE {\bfseries Input:} Relational event matrix $Y$, number of blocks $k$
    \STATE {\bfseries Result:} Estimated block assignments $\hat{C}$ and CHIP model parameters ($\hat{\vec{\pi}}, \hat{\mu}, \hat{\alpha}, \hat{\beta}$)
  \end{algorithmic}
  
  \begin{algorithmic}[1]
    \FORALL{node pairs $i \neq j$}
      \STATE $N_{ij} = $ number of events from $i$ to $j$ in $Y$
    \ENDFOR

    \STATE $\hat{C} \leftarrow \text{Spectral clustering}(N,k)$ \label{line:spectral}
    
    \FORALL{block pairs $(a,b)$} \label{line:estimationForLoop}
      \STATE Compute estimates $(\hat{m}_{ab}, \hat{\mu}_{ab})$ using \eqref{count_est}
      
      \STATE $\hat{\beta}_{ab} \leftarrow$ maximize  log-likelihood by line search
      
      \STATE $\hat{\alpha}_{ab} \leftarrow \hat{\beta}_{ab} \hat{m}_{ab}$
    \ENDFOR
    
    \STATE $\hat{\vec{\pi}} \leftarrow$ proportion of nodes in each block
    
    \STATE \textbf{return} $[\hat{C},\hat{\vec{\pi}},\hat{\mu},\hat{\alpha},\hat{\beta}]$
  \end{algorithmic}
\end{algorithm}

As with many other block models, the maximum-likelihood estimator for the discrete community assignments $C$ is intractable except for extremely small networks (e.g.~20 nodes).
We propose a scalable estimation procedure for the CHIP model that has two components as shown in Algorithm \ref{alg:estimation}: community detection and parameter estimation. 
For the community detection component, we use spectral clustering on the weighted adjacency or count matrix $N(T)$ or simply $N$ with entries $N_{ij}(T)$. 
Since this is a directed adjacency matrix, we use singular vectors rather than eigenvectors for spectral clustering (see Algorithm \ref{supp-alg:spectralClustering} in the supplementary material for details).

For the parameter estimation component, we first consider estimating the Hawkes process parameters $(\mu_{ab}, \alpha_{ab}, \beta_{ab})$ for each block pair $(a,b)$ using only the count matrix $N$, which discards event timestamps. 
Even without the event timestamps, we are able to estimate $\mu_{ab}$ and the ratio $m_{ab}=\alpha_{ab}/\beta_{ab}$, but not the parameters $\alpha_{ab}$ and $\beta_{ab}$ separately. 
Define the following terms, which are the sample mean and (unbiased) sample variance of the pairwise event counts within each block pair:
\begin{equation}
\label{eq:sample_mean_var}
\bar{N}_{ab} =\frac{1}{n_{ab}} \sum_{i,j: C_{ia}=1,C_{jb}=1}N_{ij}, \quad
S^2_{ab} = \frac{1}{n_{ab}-1} \sum_{i,j: C_{ia}=1,C_{jb}=1} (N_{ij} -\bar{N}_{ab})^2,
\end{equation}
where $n_{ab}$ denotes the number of node pairs in block pair $(a,b)$ and is given by $ n_{ab} =|a||b|$ for $ a \neq b $ and $ n_{ab} =|a||a-1|$ for $a = b$, with $|a|$ denoting the number of nodes in block $a$. 
$\bar{N}_{ab}$ and  $S^2_{ab}$ are unbiased estimators of the mean and variance, respectively, of the counts of the number of events between all node pairs $(i,j)$ in block pair $(a,b)$. 
Using $\bar{N}_{ab}$ and  $S^2_{ab}$, we propose the following method of moments estimators (conditioned on the estimated blocks) for $m_{ab}$ and $\mu_{ab}$ from the count matrix $N$:
\begin{equation}
\hat{m}_{ab} = 1-\sqrt{\frac{\bar{N}_{ab}}{S^2_{ab}}}, \quad
\hat{\mu}_{ab} = \frac{1}{T} \sqrt{\frac{(\bar{N}_{ab})^3}{S^2_{ab}}}.
\label{count_est}
\end{equation}
Finally, the vector of block assignment probabilities $\vec{\pi}$ can be easily estimated using the proportion of nodes in each block, i.e.~
$\hat{\pi}_a = \frac{1}{n} \sum_{i=1}^n \hat{C}_{ia}$ for all $a = 1, \ldots, k$.

In some prior work, exponential Hawkes processes are parameterized only in terms of $m$ and $\mu$, with $\beta$ treated as a known parameter that is not estimated \citep{zhou2013learning2,bacry2016mean,bacry2017tick}. 
In this case, the estimation procedure is complete. 
On the other hand, if we want to estimate the values of both $\alpha$ and $\beta$ rather than just their ratio, we have to use the actual event matrix $Y$ with the event timestamps. 
To separately estimate the $\alpha_{ab}$ and $\beta_{ab}$ parameters, we replace $\alpha_{ab} = \beta_{ab} m_{ab}$ in the exponential Hawkes log-likelihood for block pair $(a,b)$ then plug in our estimate $\hat{m}_{ab}$ for $m_{ab}$. Then the log-likelihood  is purely a function of $\beta_{ab}$ and can be maximized using a standard scalar optimization or line search method. 

\subsection{Selection of the Number of Blocks}
The estimation procedure in Algorithm \ref{alg:estimation} assumes that the number of blocks $k$ is provided. 
In many practical settings, $k$ is unknown, and choosing $k$ becomes a model selection problem. 
Given that CHIP uses spectral clustering on the weighted adjacency matrix $N$, model selection approaches for static block models can be used to find the optimal $k$. 
These range in complexity from the eigengap heuristic \citep{von2007tutorial} to more sophisticated methods including using eigenvalues of the non-backtracking matrix and Bethe Hessian matrix \cite{le2015estimating} and network cross validation \cite{chen2018network, li2020network}. 
Another approach, specific to the timestamped network setting we consider in this paper, is to hold out a portion of the events, e.g.~the last $20\%$, and select the $k$ that maximizes the log-likelihood on the held-out events.

\section{Theoretical Analysis of Estimators}
\label{sec:analysis}

\subsection{Analysis of Estimated Community Assignments}
\label{sec:analysis_comm}
We define the error of community detection as the misclustering error rate
$r = \inf_{\Pi} \frac{1}{n}  \sum_{i=1}^{n} 1( c_i \neq \Pi(\hat{c}_i))$, where $\Pi(\cdot)$ denotes the set of all permutations of the community labels.
Our proposed CHIP model considers directed events; however, we analyze community detection on undirected networks to better match up with the literature on analysis of spectral clustering for the SBM. 
The bounds and consistency properties we derive still apply to the directed case with only a change in the constants.  
We assume that $T \rightarrow \infty$, which can be achieved by  rescaling the time unit for event times. Under this assumption, the mean and variance of the number of events between nodes $(i,j)$ are \citep{hawkes1974cluster,hawkes1971point,lewis1969asymptotic}
\begin{equation}
    \label{eq:asy_mean}
    \nu_{ab} =  \frac{\mu_{ab}T}{1-\alpha_{ab}/\beta_{ab}}, \quad 
    \sigma^2_{ab} =  \frac{\mu_{ab}T}{(1-\alpha_{ab}/\beta_{ab})^3}.
\end{equation}

We analyze community detection error in a simplified special case of our CHIP model which is in similar spirit to a commonly-employed case in the stochastic block models literature \citep{rcy11,lei2015consistency,paul2017spectral,chin2015stochastic,gao2017achieving}. 
We provide analogous results for the general CHIP model in Section \ref{supp-sec:estimatedCommunityAssignments} of the supplementary material.
In this special case, all communities have roughly equal number of elements $|a| \asymp n/k$, all intra-community processes (diagonal block pairs) have the same set of parameters $\mu_1,\alpha_1,\beta_1$ and all inter-community processes (off-diagonal block pairs) have the same set of parameters $\mu_2,\alpha_2,\beta_2$. 
We use the notation $Y \sim \text{CHIP}(C,n,k,\mu_1,\alpha_1,\beta_1,\mu_2,\alpha_2,\beta_2)$
to denote a relational event matrix $Y$ generated from this simplified model. 
Define $m_1 = \alpha_1/\beta_1$ and $m_2 = \alpha_2/\beta_2$. Let $\nu_1 = \mu_1/(1-m_1)$ and $\nu_2 = \mu_2/(1-m_2)$, while $\sigma^2_1 = \mu_1/(1-m_1)^3$ and $\sigma^2_2 = \mu_2/(1-m_2)^3$. Assume $\nu_1>\nu_2$, $\nu_1 \asymp \nu_2$, and $\sigma_1 \asymp \sigma_2$, where the asymptotic equivalence is with respect to both $n$ and $T$. 
These assumptions imply that the expected number of events are higher between two nodes in the same  community compared to two nodes in different communities and that the asymptotic dependence on $n$ and $T$ are the same for both set of parameters. This setting is useful to understand detectability limits and has been widely employed in the literature on stochastic block models 
\citep{abbe2015community,gao2017achieving,chin2015stochastic,vu2018simple,abbe2017community,paul2017spectral}. 
In this setting, we have the following upper bound on the misclustering error rate.

\begin{thm}
\label{cor:misclusWeighted}
Let $Y \sim \text{CHIP}(C,n,k,\mu_1,\alpha_1,\beta_1,\mu_2,\alpha_2,\beta_2)$. The misclustering error rate for spectral clustering on the weighted adjacency matrix $N$ at time $T \rightarrow \infty$ is
\begin{equation*}
r \lesssim \frac{T \sigma^2_1 n}{(n/k)^2(\nu_2-\nu_1)^2T^2} \asymp  \frac{k^2}{nT}\frac{\sigma^2_1 }{(\nu_1-\nu_2)^2}.
\end{equation*}
\end{thm}
We note that if the set of parameters $\mu,\alpha,\beta$ remain constant as a function of $n$ and $T$ then the misclustering error rate decreases as $1/T$ with increasing $T$, decreases as $1/n$ with increasing $n$, and increases as $k^2$ with increasing $k$. Hence, as we observe the process for more time, spectral clustering on $N$ has lower error rate. The rate of convergence with increasing $T$ is the same as one would obtain for detecting an average community structure if discrete snapshots of the network were available over time \citep{pensky2019spectral,paul2017spectral,bhattacharyya2018spectral}. The dependence of the misclustering error rate on $n$ and $k$ is what one would expect from the SBM literature. 

Theorem \ref{cor:misclusWeighted} applies also in the sparse graph setting. 
We let  $\mu \asymp 1/[f(n)g(T)]$, a function of $n$ and $T$, and explore various sparsity settings by varying $f$ and $g$ in Section \ref{supp-sec:SpecialCase} of the supplementary material. 
Our proofs allow $\mu$ to vary with $n$ and $T$ and can be as small as $\log(n) / (nT)$.
A key component in the proof of Theorem \ref{cor:misclusWeighted} is a bound from \citet{bandeira2016sharp}. 
In Section \ref{supp-sec:estimatedCommunityAssignments} of the supplementary material, we provide the proof of Theorem \ref{cor:misclusWeighted}, an analogous theorem for the general CHIP model, as well as theorems for spectral clustering on an unweighted adjacency matrix.

\subsection{Analysis of Estimated Hawkes Process Parameters} \label{sec:HawkesParamEstimators}
As discussed in Section \ref{sec:EstimationProcedure}, we are able to estimate $m = \alpha/\beta$ and $\mu$ from the count matrix $N$ using \eqref{count_est}. 
We analyze these estimators assuming a growing number of nodes $n$ and time duration $T$. 
We do not put any assumption on the distribution of the counts; we only require that $T$ is large enough such that the asymptotic mean and variance equations in \eqref{eq:asy_mean} hold.
The sample mean $\bar{N}_{ab}$ and sample variance $S^2_{ab}$ of the counts are unbiased estimators of $\nu_{ab}$ and $\sigma^2_{ab}$, respectively. 
The following theorem shows that these estimators are consistent and asymptotically normal.

\begin{thm}
\label{hawkes_thm}
Define $n_{\min} =\min_{a,b} n_{ab}$. The estimators for $m_{ab}$ and $\mu_{ab}$ have the following asymptotic distributions as $n_{\min} \rightarrow \infty$ and $T \rightarrow \infty$: 
\[
\sqrt{n_{ab}} \left(\hat{m}_{ab} - \left(1-\sqrt{\frac{\nu_{ab}}{\sigma^2_{ab}}}\right)\right) \overset{d}{\rightarrow} \mathcal{N} \left(0,\frac{1}{4\nu_{ab}}\right), \;
\sqrt{n_{ab}}\left(\hat{\mu}_{ab}T - \frac{(\nu_{ab})^{3/2}}{\sigma_{ab}}\right) \overset{d}{\rightarrow} \mathcal{N} \left(0,\frac{9}{4}\nu_{ab}\right).
\]
\end{thm}
Using Theorem \ref{hawkes_thm}, we obtain confidence intervals for $\mu$ and $m$, in Section \ref{supp-sec:HawkesParamConfInterval} of the supplementary material.
In the simplified special case of Theorem \ref{cor:misclusWeighted}, we have equal community sizes so $n_{ab} \asymp (n/k)^2$. 
Therefore, the condition $n_{\min} \rightarrow \infty$ boils down to $(n/k)^2 \rightarrow \infty$, which is a reasonable assumption. 
Theorem \ref{hawkes_thm} guarantees convergence of our estimators for $\mu$ and $m$ with the asymptotic mean-squared errors (MSEs) decreasing at the rate $n_{ab} \asymp (n/k)^2$ under the assumption that the community structure is correctly estimated. 
Next, we provide an ``end-to-end'' guarantee for the convergence of the asymptotic MSE to 0 for estimating the mean number of events in each block pair $\nu_{ab}$ using the sample mean $\bar{N}_{ab}$ incorporating the error in estimating communities using spectral clustering from Theorem \ref{cor:misclusWeighted}. 
\begin{thm}
\label{endtoend}
Assume $n_{ab} \asymp (n/k)^2$. 
The weighted average of asymptotic MSEs in estimating $\nu_{ab}$ using the estimator $\bar{N}_{ab}$ with communities estimated by spectral clustering is
\[
\frac{\sum_{ab} n_{ab} E[(\bar{N}_{ab} - \nu_{ab})^2]}  {\sum_{ab} n_{ab}} \lesssim \frac{kT}{n} \max\left\{\sigma_1^2, \frac{k^2\sigma_1^2\nu_2^2}{(\nu_1-\nu_2)^2} \right\}.
\]
For comparison, under the assumption that the community structure is correctly estimated, the weighted average of asymptotic MSEs in estimating $\nu_{ab}$ using the estimator $\bar{N}_{ab}$ is
\[
\frac{\sum_{ab} n_{ab} E[(\bar{N}_{ab} - \nu_{ab})^2]} {\sum_{ab} n_{ab}}  =  \frac{k^2T\sigma_1^2}{n^2}.
\]

\end{thm}
Theorem \ref{endtoend} guarantees that the MSE for estimating Hawkes process parameters decreases at least at a linear rate with increasing $(n/k)$ when the error from community detection is taken into account instead of the quadratic rate when the error is not taken into account.
The proofs for Theorems \ref{hawkes_thm} and \ref{endtoend} are provided in Section \ref{supp-sec:estimatedHawkesParamProofs} of the supplementary material.

\section{Experiments}

We begin with a set of simulation experiments to assess the accuracy of our proposed estimation procedure and verify our theoretical analysis. 
We then present several experiments on real data involving both prediction and model-based exploratory analysis. 
Additional experiments are provided in Section \ref{supp-sec:additionalExperiments} of the supplementary material, along with the code\footnote{Code available on GitHub: \url{https://github.com/IdeasLabUT/CHIP-Network-Model}.} to replicate all experiments.

\subsection{Community Detection on Simulated Networks with Varying \texorpdfstring{$T$}{T}, \texorpdfstring{$n$}{n}, and \texorpdfstring{$k$}{k}}

\begin{figure*}[t]
    \centering
    \subfigure[Fixed $n=256$]{
    \includegraphics[width=1.7in]{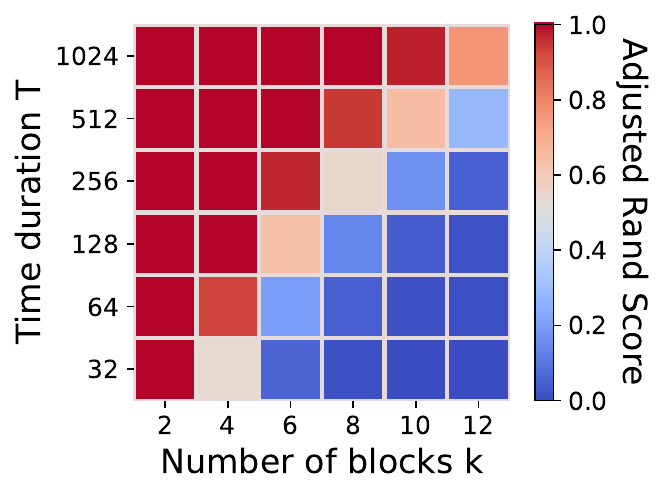}
    \label{fig:commDetectionFixedN}
    }
    \subfigure[Fixed $T=64$]{
    \includegraphics[width=1.7in]{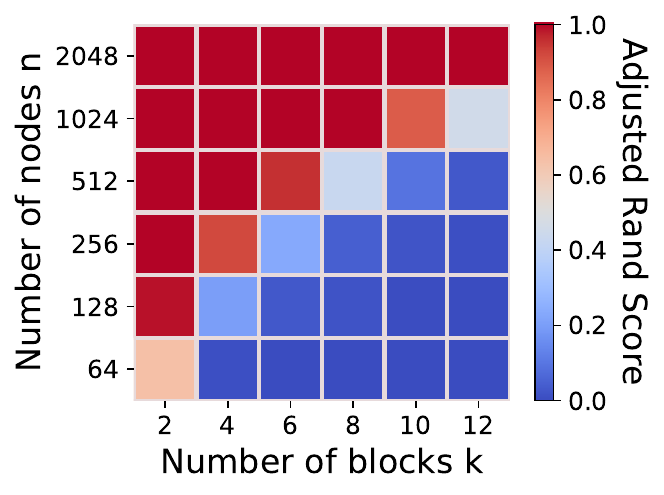}
    \label{fig:commDetectionFixedT}
    }
    \subfigure[Fixed $k=8$]{
    \includegraphics[width=1.7in]{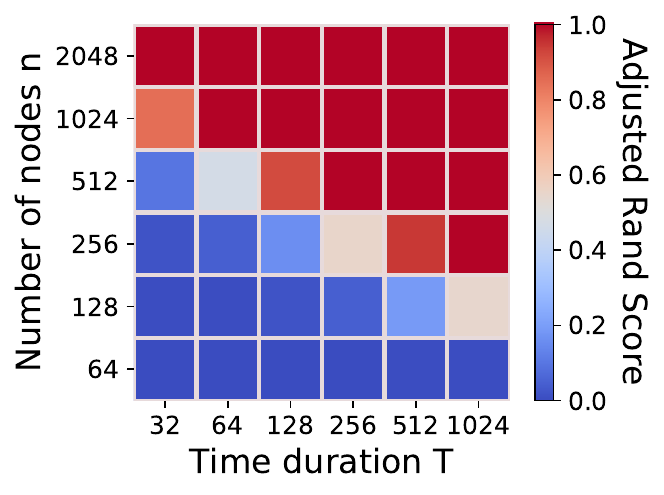}
    \label{fig:commDetectionFixedK}
    }
    \caption{Heat map of adjusted Rand score of spectral clustering on weighted adjacency matrix, with varying $T$, $n$, and $k$, averaged over 30 simulated networks.}
\label{fig:commDetectionHeatMap}
\end{figure*}

We simulate networks from the simplified CHIP model while varying two of $T$, $n$, and $k$ simultaneously. 
We choose parameters $\mu_1 = 0.085$, $\mu_2 = 0.065$, $\alpha_1 = \alpha_2 = 0.06$, and $\beta_1 = \beta_2 = 0.08$. 
The upper bounds on the error rates in Theorem \ref{cor:misclusWeighted} involve all three parameters $n,k,T$ simultaneously, making it difficult to interpret the result. To better observe the effects of $n,k,T$ and their relationship with respect to each other, we perform three separate simulations each time varying two and fixing the other one. 
The community detection accuracy averaged over 30 simulations using the weighted adjacency matrix $N$ as two of $T$, $n$, and $k$ are varied is shown in Figure \ref{fig:commDetectionHeatMap}. 
Since the estimated community assignments will be permuted compared to the actual community labels, we evaluate the community detection accuracy using the adjusted Rand score \citep{hubert85}, which is $1$ for perfect community detection and has an expectation of $0$ for a random assignment.

Note that Theorem \ref{cor:misclusWeighted} predicts that the misclustering error rate varies as $k^2/(nT)$ if all three parameters are varied. Figure \ref{fig:commDetectionFixedN} shows the accuracy to be low for small $T$ and large $k$. As we simultaneously increase $T$ and decrease $k$ the accuracy improves until the adjusted Rand score reaches $1$. We also note that it is possible to obtain high accuracy either with increasing $T$ or decreasing $k$ or with both even when $n$ is fixed. This is in line with the prediction from Theorem \ref{cor:misclusWeighted} that the misclustering error rate varies as $k^2/T$ if $n$ remains fixed. We observe a similar effect of increasing accuracy with increasing $n$ and decreasing $k$ when $T$ is kept fixed in Figure \ref{fig:commDetectionFixedT}. Finally, Figure \ref{fig:commDetectionFixedK} verifies the prediction that accuracy increases with both increasing $n$ and $T$ for a fixed $k$.

\subsection{Hawkes Process Parameter Estimation on Simulated Networks} \label{sec:consistentParam_est}

Next, we examine the estimation accuracy of the Hawkes process parameter estimates as described in Section \ref{sec:HawkesParamEstimators}. 
We simulate networks from the simplified CHIP model with $k=4$ blocks, duration $T=10,\!000$ and parameters $\mu_1 = 0.0011$, $\mu_2 = 0.0010$, $\alpha_1 = 0.11$, $\alpha_2 = 0.09$, $\beta_1 = 0.14$, and $\beta_2 = 0.16$ so that each parameter is different between block pairs. 
We then run the CHIP estimation procedure: spectral clustering followed by Hawkes process parameter estimation.

\begin{figure}[t]
    \centering
    \subfigure[$\mu$: base intensity (MSE decay rate: 2.00)]{
    \includegraphics[width=1.2in]{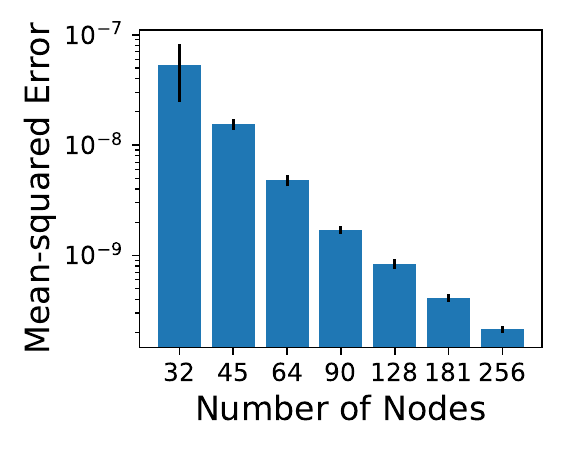}
    \label{fig:consMseMu}
    } \;
    \subfigure[$m$: $\alpha$ to $\beta$ ratio (MSE decay rate: 2.05)]{
    \includegraphics[width=1.2in]{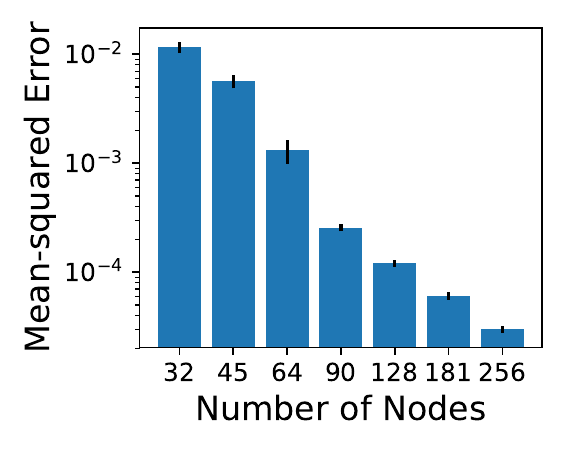}
    \label{fig:consMseM}
    } \;
    \subfigure[$\alpha$: intensity jump size (MSE decay rate: 2.01)]{
    \includegraphics[width=1.2in]{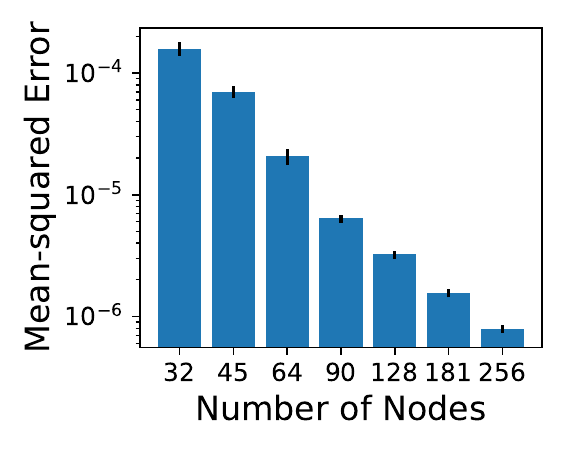}
    \label{fig:consMseAlpha}
    } \;
    \subfigure[$\beta$: intensity decay rate (MSE decay rate: 2.08)]{
    \includegraphics[width=1.2in]{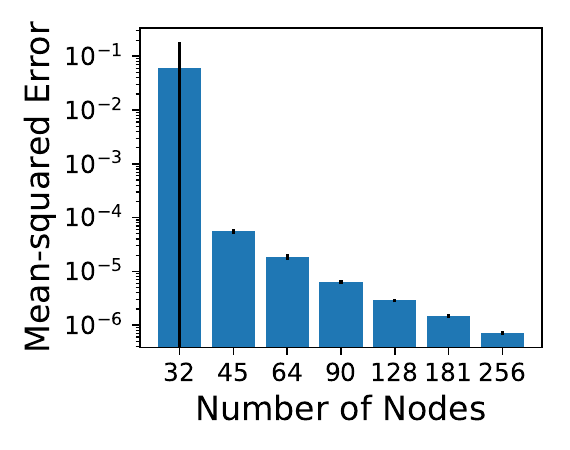}
    \label{fig:consMseBeta}
    }
    \caption{Mean-squared errors (MSEs) of CHIP's Hawkes parameter estimators averaged over 100 simulations ($\pm$ 2 standard errors) on a log-log plot. MSEs for all four parameters decreases as the number of nodes increases, with the estimated decay rate (exponent) beginning at 90 nodes listed.}

\label{fig:consMse}
\end{figure}

Figure \ref{fig:consMse} shows the mean-squared errors (MSEs) of all four estimators decay quadratically as $n$ increases. 
Theorem \ref{hawkes_thm} states that $\hat{m}$ and $\hat{\mu}$ are consistent estimators with MSE decreasing at a quadratic rate for growing $n$ with known communities. 
Here, we observe the quadratic decay even with communities estimated by spectral clustering, where the mean adjusted Rand score is increasing from $0.6$ to $1$ as $n$ grows. 
We observe that $\alpha$ and $\beta$ are also accurately estimated for growing $n$ even though $\beta$ is estimated using a line search for which we have no guarantees.

\subsection{Comparison with Other Models on Real Networks} \label{sec:realNetworksModelComparison}

We perform experiments on three real network datasets.
Each dataset consists of a set of events where each event is denoted by a sender, a receiver, and a timestamp. The  MIT Reality Mining \citep{eagle2009inferring} and Enron \citep{klimt2004enron} datasets were loaded and preprocessed identically to \citet{Dubois2013} to allow for a fair comparison with their reported values.
On the Facebook wall posts dataset  \citep{viswanath2009evolution}, we use the largest connected component of the network excluding self loops ($43,953$ nodes).
Additional details about the datasets and preprocessing are provided in Section \ref{supp-sec:datasetDescriptions} of the supplementary material.

\begin{table}[t]
    \renewcommand{\arraystretch}{1.1}
	\centering
    \caption{Mean test log-likelihood per event for each real network dataset across all models. Larger (less negative) values indicate better predictive ability. 
    Bold entry denotes best fit for a dataset.
    Results for REM are reported values from \citet{Dubois2013}. 
    Poisson denotes spectral clustering followed by estimating a Poisson process baseline model.
    \textsuperscript{*}The BHM local search does not scale up to the Facebook network, so we report results using the (less accurate) spectral clustering procedure.}
    \label{tab:realDataModelComparison}
	\begin{tabular}{cccccccc}
	\hline
	Dataset & Statistics & Model & $k=1$ & $k=2$ & $k=3$ & $k=10$ & Best $k$ \\
	\hline
	\multirow{4}{*}{Reality} 
	& \multirow{4}{*}{\makecell{$n=70$\\$l_{\text{train}} = 1,\!500$\\$l_{\text{test}} = 661$}}
    & CHIP & $-4.83$          & $-4.88$ & $-5.06$ & $-6.69$ & $\mathbf{-4.83}$ ($k=1$) \\
	& & REM  & $-6.78$          & $-7.42$ & $-6.11$ & $-6.61$ & $-6.11$ ($k=3$) \\
	& & BHM  & $-9.05$          & $-7.56$ & $-7.60$ & $-5.74$ & $-5.37$ ($k=50$)\\
	& & Poisson & $-10.3$ & $-10.4$ & $-9.63$ & $-9.38$ & $-8.51$ ($k=32$)\\
	\hline
	\multirow{4}{*}{Enron}
	& \multirow{4}{*}{\makecell{$n=142$\\$l_{\text{train}} = 3,\!000$\\$l_{\text{test}} = 1,\!000$}}
    & CHIP & $-5.63$          & $-5.61$ & $-5.65$ & $-7.15$ & $\mathbf{-5.61}$ ($k=2$) \\
	& & REM  & $-7.02$          & $-6.86$ & $-6.84$ & $-7.26$ & $-6.84$ ($k=3$) \\
	& & BHM  & $-8.72$          & $-8.43$ & $-8.39$ & $-7.93$ & $-7.49$ ($k=8$) \\
	& & Poisson & $-11.9$ & $-11.4$ & $-11.5$ & $-12.0$ & $-11.4$ ($k=4$)\\
	\hline
	\multirow{3}{*}{Facebook} 
	& \multirow{3}{*}{\makecell{$n=43,\!953$\\$l_{\text{train}} = 682,\!266$\\$l_{\text{test}} = 170,\!567$}}
    & CHIP & $-9.54$ & $-9.58$ & $-9.58$ & $-9.61$ & $\mathbf{-9.46}$ ($k=9)$ \\
	& & BHM\textsuperscript{*}  & $-16.0$ & $-15.7$ & $-16.2$ & $-14.7$ & $-14.4$ ($k=22$)\\
	& & Poisson & $-20.8$ & $-21.1$ & $-21.1$ & $-20.6$ & $-19.2$ ($k=55$)\\
	\hline
	\end{tabular}

\end{table}

We fit our proposed Community Hawkes Independent Pairs (CHIP) model as well as the Block Hawkes Model (BHM) \citep{junuthula2017block} to all three real datasets and evaluate their fit. 
We also compare against a simpler baseline: spectral clustering with a homogeneous Poisson process for each node pair.
For each model, we also compare against the case $k=1$, where no community detection is being performed. 
We do not have ground truth community labels for these real datasets so we cannot evaluate community detection accuracy. 
Instead, we use the mean test log-likelihood per event as the evaluation metric, which allows us to compare against the reported results in \citet{Dubois2013} for the relational event model (REM). 
Since the log-likelihood is computed on the test data, this is a measure of the model's ability to \emph{forecast future events} rather than detect communities.

As shown in Table \ref{tab:realDataModelComparison}, CHIP outperforms all other models in all three datasets. 
Note that test log-likelihood is maximized for CHIP at relatively small values of $k$ compared to the BHM. 
This is because CHIP assumes independent node pairs whereas the BHM assumes all node pairs in a block pair are dependent.
Thus, the BHM needs a higher value for $k$ in order to model independence. 
This difference is particularly visible for the Reality Mining data, where CHIP with $k=1$ is the best predictor of the test data, while the best BHM has $k=50$ on a network with only $70$ nodes!
These both suggest a weak community structure that is not predictive of future events in the Reality Mining data, whereas community structure does appear to be predictive in  the Enron and Facebook data.

\begin{wrapfigure}[12]{r}{1.8in}
\centering
\vspace{-12pt}
\includegraphics[width=1.8in]{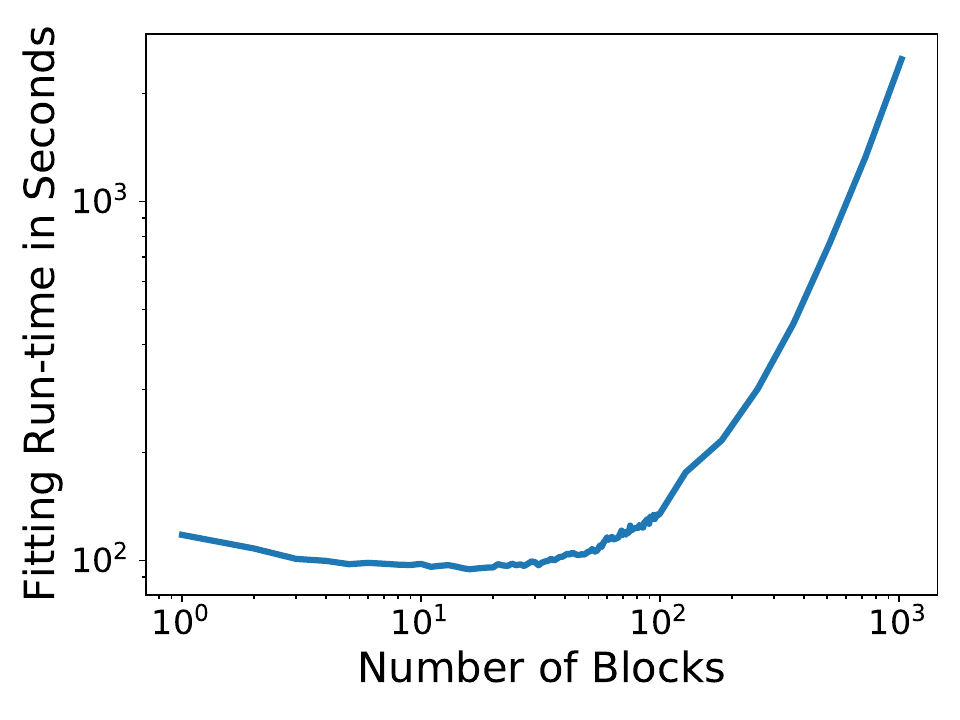}
\vspace{-11pt}
\caption{CHIP's fitting runtime on the Facebook data on a log-log scale with increasing $k$.}
\label{fig:chip-fb-runtime-k}
\end{wrapfigure}

In addition to the improved predictive ability of CHIP compared to the BHM, the computational demand is also significantly decreased. 
Fitting the CHIP model for each value of $k$ took on average $0.15$ s and $0.3$ s on the Reality Mining and Enron datasets, respectively, while the BHM took on average $250$ s and $30$ m, mostly due to the time-consuming local search\footnote{\label{footnote:workstaionProcessingPower}Experiments were run on a workstation with 2 Intel Xeon 2.3 GHz CPUs with a total of 36 cores.}.
We did not implement the MCMC-based inference procedure for the REM and thus do not have results for REM on the Facebook data or computation times. 
The approach of holding out a test set of events and evaluating test log-likelihood can also be used for selection of the number of blocks $k$. 
As shown in Figure \ref{fig:chip-fb-runtime-k}, on the Facebook data, there is hardly any increase in the runtime of CHIP for $k < 100$, and it is manageable even for $k=1,\!000$.

\subsection{Model-Based Exploratory Analysis of Facebook Wall Post Network} \label{sec:fbExploratoryAnalysis}

\begin{figure}[t]
    \centering
    \subfigure[Adjacency matrix with blocks]{
    \includegraphics[height=1.3in]{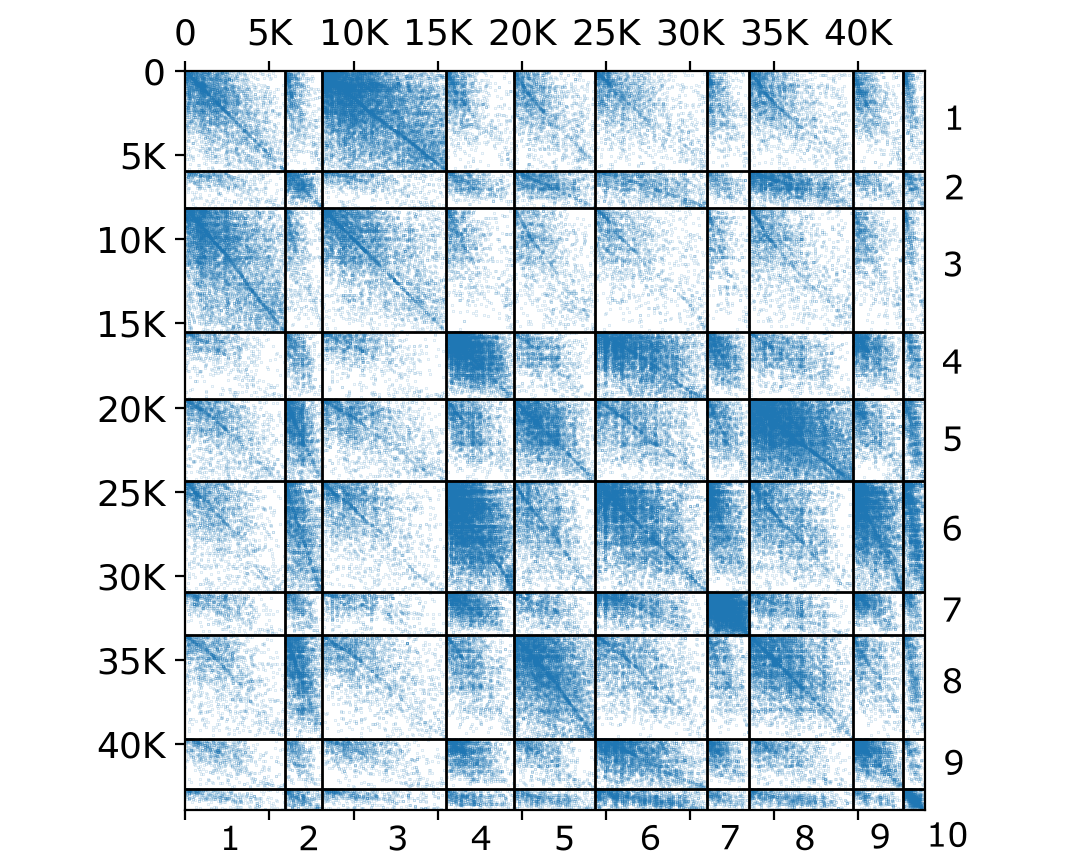}
    \label{fig:fbChpAdjMat}
    }
    \subfigure[$\mu$: base intensity]{
    \includegraphics[height=1.3in]{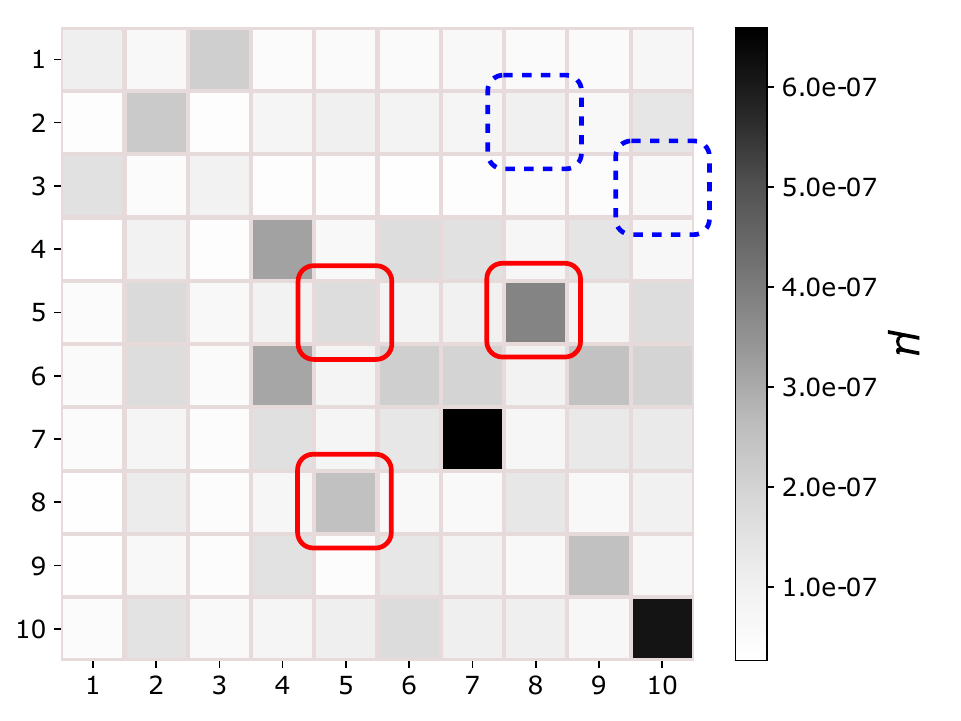}
    \label{fig:fbChpMu}
    }
    \subfigure[$m$: branching ratio $\alpha/\beta$]{
    \includegraphics[height=1.3in]{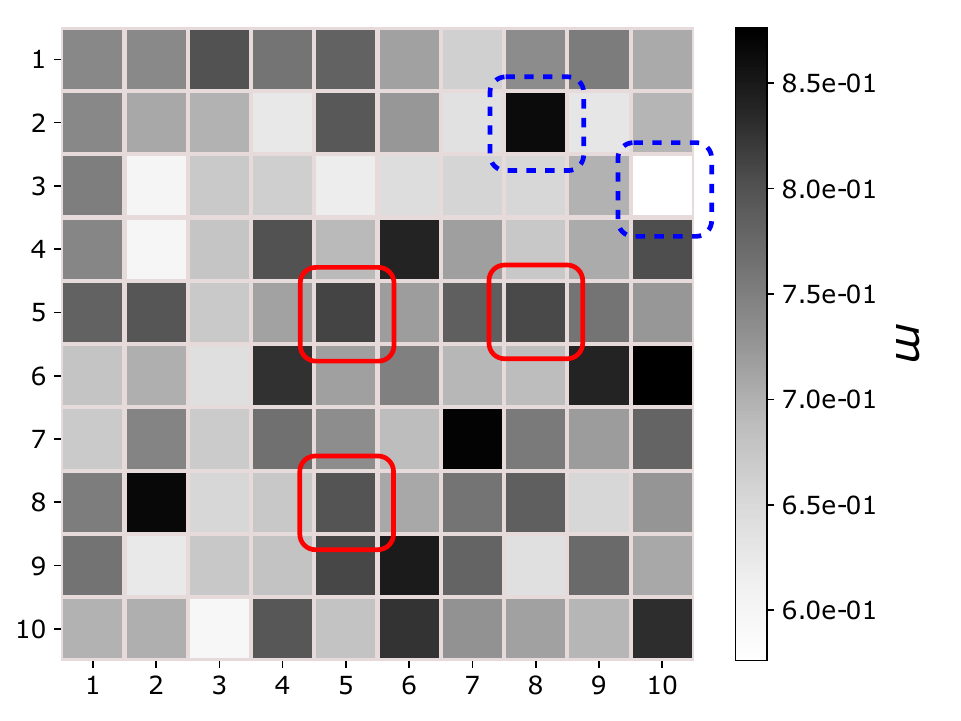}
    \label{fig:fbChpM}
    }
    \caption{Inferred CHIP parameters on the largest connected component of the Facebook Wall Posts dataset with $k=10$. Axis labels denote block numbers. Each tile corresponds to a block pair where $(a,b)$ denotes row $a$ and column $b$. 
    Boxed block pairs in the heatmap are discussed in the body text.}
\label{fig:fbChpFit}
\end{figure}

We use CHIP to perform model-based exploratory analysis to understand the behavior of different groups of users in the Facebook wall post network. 
We consider all $852,\!833$ events and choose $k=10$ blocks using the eigengap heuristic \citep{von2007tutorial}, which required $141$ s to fit. 
Note that the CHIP estimation procedure can scale up to a much higher number of communities also---fitting CHIP to the Facebook data with $k=1,\!000$ communities took just under $50$ minutes! 
The adjacency matrix permuted by the block structure is shown in Figure \ref{fig:fbChpAdjMat}, and heatmaps of the fitted CHIP parameters are shown in Figures \ref{fig:fbChpMu} and \ref{fig:fbChpM}. 
Diagonal block pairs on average have a base intensity $\mu$ of $2.8 \times 10^{-7}$, which is higher compared to $9.5 \times 10^{-8}$ for off-diagonal block pairs, indicating an underlying assortative community structure. 
However, not all blocks have higher rates of within-block posts, e.g.~$\mu_{5,8} > \mu_{5,5}$ and $\mu_{8,5} > \mu_{5,5}$, as shown in red boxes in Figure \ref{fig:fbChpMu},  
which illustrates that the CHIP model does not discourage inter-block events. These patterns often occur in social networks, for instance, if there are communities with opposite views on a particular subject.

While the structure of $\mu$ reveals insights on the baseline rates of events between block pairs, the structure of the branching ratio $m = \alpha/\beta$ shown in Figure \ref{fig:fbChpM} reveals insights on the burstiness of events between block pairs. 
For some block pairs, such as $(3,10)$, there are very low values of $\alpha$ and $\beta$ indicating the events are closely approximated by a homogeneous Poisson process, while some block pairs such as $(2,8)$ are extremely bursty despite low baseline rates. Both block pairs are shown in blue dashed boxes.
The different levels of burstiness of block pairs cannot be seen from aggregate statistics such as the the count matrix $N$.

\section{Conclusion}

We introduced the Community Hawkes Independent Pairs (CHIP) model for timestamped relational event data. The CHIP model has many similarities with the Block Hawkes Model (BHM) \citep{junuthula2017block}; however, in the CHIP model, events among any two node pairs are independent, which enables both tractable theoretical analysis and scalable estimation. 
We demonstrated that an estimation procedure using spectral clustering followed by Hawkes process parameter estimation provides consistent estimates of the communities and Hawkes process parameters for a growing number of nodes and time duration.
Lastly, we showed that CHIP also provides better fits to several real networks compared to the Relational Event Model \citep{Dubois2013} and the BHM. 
It also scales to considerably larger data sets, including a Facebook wall post network with over $40,\!000$ nodes and $800,\!000$ events.

There are several limitations to the CHIP model and our proposed estimation procedure. 
Assuming all node pairs to have independent Hawkes processes simplifies analysis and increases scalability but also reduces the flexibility of the model compared to multivariate Hawkes process-based models that specifically encourage reciprocity \citep{Blundell2012, miscouridou2018modelling}. 
Additionally, our estimation procedure uses unregularized spectral clustering to match our theoretical analysis in Section \ref{sec:analysis}. 
We note that regularized versions of spectral clustering \citep{chaudhuri2012spectral, amini2013pseudo, qin13, joseph2016impact, zhang2018understanding} have been found to perform better empirically and would likely improve the community detection accuracy in the CHIP model. 
Methods that jointly estimate the community structure and Hawkes process parameters, such as the local search and variational inference approaches explored in \citet{junuthula2017block} for the Block Hawkes Model could also improve estimation accuracy of both. 
Also, methods that integrate change point detection with estimation for continuous-time block models could be used to allow for community structure to change over time \citep{corneli2018multiple}, resulting in more flexible models.

\section*{Broader Impact}
Our proposed CHIP model can be applied to analyze any type of timestamped relational event data. 
In this paper, we considered analysis of mobile phone calls, emails, and user interactions on on-line social networks. 
However, timestamped relational event data is used in a variety of other disciplines, including financial mathematics, e.g.~transactions between traders in financial markets \citep{bacry2015hawkes}; political science, e.g.~military deployments between countries \citep{maoz2019dyadic,Blundell2012}; and sociology, e.g.~homicides between gangs in a city \citep{papachristos2009murder,Linderman2014}.
Thus, our CHIP model can have broader impact to society through the advancement of multiple research disciplines. 

The CHIP model, like other generative models for dynamic networks, can be used for forecasting, e.g.~to predict which nodes are likely to have an event, as well as the number of events during a specified time interval. 
For some applications, the forecasts may themselves be used to affect decision making. 
For example, in public policy, crime forecasting can be used for predictive policing, which affects the allocation of police resources to different locations over time. 
This can have societal benefits, as a recent randomized controlled field trial for predictive policing using Hawkes process models for prediction demonstrated a 7\% reduction in crime \citep{mohler2015randomized}, but also potential for negative consequences like arrests that are biased with respect to minority communities, although such consequences were not observed in the randomized trial \citep{brantingham2018does}. 
In this paper, we analyzed a publicly available anonymized Facebook on-line social network dataset, so we are not aware of negative consequences that may result from our proposed model.

\begin{ack}
This material is based upon work supported by the National Science Foundation grants DMS-1830412 and IIS-1755824.
\end{ack}

\bibliography{references}
\bibliographystyle{unsrtnat}

\makeatletter\@input{xxsupp.tex}\makeatother
\end{document}


\date{}

\maketitle

\appendix

\section{Additional Details on Estimation Procedure}

\subsection{Community Detection}
The spectral clustering algorithm for directed networks that we consider in this paper is shown in Algorithm \ref{alg:spectralClustering}. 
It can be applied either to the weighted adjacency (count) matrix $N$ or the unweighted adjacency matrix $A$, where $A_{ij} = 1\{N_{ij} > 0\}$ and $1\{\cdot\}$ denotes the indicator function of the argument. 
This algorithm is used for the community detection step in our proposed CHIP estimation procedure. 
For undirected networks, which we use for the theoretical analysis in Section \ref{sec:analysis}, spectral clustering is performed by running k-means clustering on the rows of the \emph{eigenvector} matrix of $N$ or $A$, not the rows of the concatenated singular vector matrix.

\begin{algorithm}[t]
  \caption{Spectral clustering algorithm for community detection in directed networks}
  \label{alg:spectralClustering}
  \begin{algorithmic}
  \STATE {\bfseries Input:} Adjacency Matrix $N$, number of blocks $k$
  \STATE {\bfseries Result:} Estimated block assignments $\hat{C}$
  \end{algorithmic}
  
  \begin{algorithmic}[1]
  \STATE Compute singular value decomposition of $N$
  \STATE $\hat{\Sigma} \leftarrow$ diagonal matrix of $k$ largest singular values of $N$
  \STATE $\hat{U}, \hat{V} \leftarrow$ left and right singular vectors of $N$ corresponding to $k$ largest singular values
  \STATE $\hat{Z} \leftarrow $ concatenate$(\hat{U}$, $\hat{V})$
  \STATE Normalize the magnitude of each row of $\hat{Z}$ to 1
  \STATE $\hat{C} \leftarrow$ k-means clustering on rows of $\hat{Z}$
  \STATE \textbf{return} $\hat{C}$
  \end{algorithmic}
\end{algorithm}

\subsection{Estimation of Hawkes process parameters}
\citet{ozaki1979maximum} derived the log-likelihood function for Hawkes processes with exponential kernels, which takes the form:
\begin{equation}
\label{eq:hawkes_log_lik}
\log \mathcal{L} = -\mu T + 
\sum_{q=1}^{l}{\frac{\alpha}{\beta} \{e^{-\beta (T - t_q)} - 1\}} + 
\sum_{q=1}^{l}{\log(\mu + \alpha w(q))}
\end{equation}
where $w(q) = \sum_{q':t_{q'} < t_{q}}{e^{-\beta (t_q - t_{q'})}}$.
Moreover, $w(q)$ can be computed recursively using $w(q) = e^{-\beta (t_q - t_{q-1})}(1 + w(q - 1))$, with the added base case of $w(1)=0$, which drops the double summation in the last term and decreases the computational complexity of the log-likelihood from $\mathcal{O}(l^2)$ to $\mathcal{O}(l)$ \citep{laub2015hawkes}.
The three parameters $\mu, \alpha, \beta$ can be estimated by maximizing \eqref{eq:hawkes_log_lik} using standard numerical methods for non-linear optimization \citep{nocedal2006numerical}.

In our CHIP model, we have separate $(\mu, \alpha, \beta)$ parameters for each block pair $(a,b)$. 
We provide closed-form equations for estimating $m_{ab} = \alpha_{ab}/\beta_{ab}$ and $\mu_{ab}$ in \eqref{count_est}.
To separately estimate the $\alpha_{ab}$ and $\beta_{ab}$ parameters, we replace $\alpha_{ab} = \beta_{ab} m_{ab}$ in the exponential Hawkes log-likelihood \eqref{eq:hawkes_log_lik} for block pair $(a,b)$ to obtain
\begin{align}
\label{log_lik_m}
\log \mathcal{L}(\beta_{ab} | C, [\vec{t}_{ij}]_{i,j=1}^n) &=  \sum_{i,j: C_{ia}=1,C_{jb}=1} \Bigg\{-\mu_{ab} T \nonumber \\
&\qquad + \sum_{q=1}^{N_{ij}}{m_{ab} \{e^{-\beta_{ab} (T - t_{ij}^q)} - 1\}}
+ \sum_{q=1}^{N_{ij}}{\log(\mu_{ab} + \beta_{ab} m_{ab} w_{ij}(q))}\Bigg\}
\end{align}
where $w_{ij}(q) = \sum_{q':t_{ij}^{q'} <t_{ij}^q}e^{-\beta_{ab}(t_{ij}^{q}-t_{ij}^{q'})}$ for $q\geq 2$ and $w_{ij}(1)=0$. 
We substitute in the estimates for $m_{ab}$ and $\mu_{ab}$ from \eqref{count_est}. 
Then the log-likelihood \eqref{log_lik_m} is purely a function of $\beta_{ab}$ and can be maximized using a standard scalar optimization or line search method. 
In our experiments, we perform the line search using SciPy's function  \texttt{minimize\_scalar(method="bounded")}.

\section{Additional Theoretical Analysis of Estimators}
We present additional results on estimation of community assignments in Section \ref{sec:estimatedCommunityAssignments} along with proofs and discussions. We then provide proofs of our results for estimated Hawkes process parameters in Section \ref{sec:proof_param} along with discussions on obtaining confidence intervals for the estimated Hawkes process parameters.

\subsection{Estimated Community Assignments} \label{sec:estimatedCommunityAssignments}
We define the notation $Y \sim \text{CHIP}(C,n,k,\mu,\alpha,\beta)$ to denote that relational event matrix $Y$ is generated from a CHIP model with $n$ nodes, $k$ blocks, community assignment matrix $C$ and Hawkes process parameter matrices $(\mu, \alpha, \beta)$. 
To characterize the misclustering rate of a spectral clustering algorithm applied to $N$, we define the following quantities. Let $\lambda_{\min}(E[N])$ denote the minimum in absolute value non-zero eigenvalue of the matrix $E[N]$. Define
\begin{align}
s &= \sqrt{T} \max_{a} \sqrt{\sum_{b} |b|\frac{\mu_{ab}}{(1-\alpha_{ab}/\beta_{ab})^3}}
\label{eq:const_s},\\
s_1 & = \sqrt{T} \max_{a,b} \sqrt{\frac{\mu_{ab}}{(1-\alpha_{ab}/\beta_{ab})^3}}.
\label{eq:const_s1}
\end{align}
Then we have the following upper bound on the misclustering error rate.
 
\begin{thm}
Let $Y \sim \text{CHIP}(C,n,k,\mu,\alpha,\beta)$. 
Then, with probability at least $1-1/n$, the misclustering error rate for spectral clustering on the weighted adjacency matrix $N$ at time $T \rightarrow \infty$ is
\begin{equation*}
r \leq 64(2+\epsilon_1) |a|_{\max} \, k \,
\frac{\left\{(1+\epsilon)(2s + \frac{6}{\log (1+\epsilon)}s_1\sqrt{\log n}) + s_1\sqrt{\log n} \right\}^2}{n(\lambda_{\min}(E[N])^2},
\end{equation*}
where $0<\epsilon<1/2$ and $\epsilon_1>0$ are constants.
\label{errorN}
\end{thm}

Theorem \ref{errorN} provides an upper bound to the error rate of spectral clustering on the weighted adjacency matrix $N$ in the setting $T \rightarrow \infty$. 
Note that the assumption of $T \rightarrow \infty$ does not preclude us from being able to analyze scenarios where the network is sparse since the expected number of events between a pair of nodes $\nu_{ab}$ can be made constant or even $o(1)$ by setting $\frac{\mu_{ab}}{1-\alpha_{ab}/\beta_{ab}}= O(1/T)$ and $\frac{\mu_{ab}}{1-\alpha_{ab}/\beta_{ab}} = o(1/T)$ respectively. 

It is also possible to obtain an expression for the mean as a function of $T$ without the assumption of $T \rightarrow \infty$ using stochastic differential equations \citep{laub2015hawkes,da2014hawkes}. In particular, if we substitute the starting intensity $\lambda_0 = \mu$, i.e., the process starts with baseline intensity as we have assumed throughout, and the starting number of events $N_0=0$, then from the result of \cite{da2014hawkes} and \cite{daw2018queues},
\[
E[N_{ij}] =  \frac{\mu_{ab}T}{1-\alpha_{ab}/\beta_{ab}} - \frac{\mu_{ab}\alpha_{ab}\left[1-e^{-(\beta_{ab}-\alpha_{ab})T}\right]}{(\beta_{ab}-\alpha_{ab})^2}.
\]
We note that there is a small negative correction term to the asymptotic mean, since $\mu_{ab}, \alpha_{ab}, \beta_{ab}, T$ are all non-negative. 
The effect of this term is negligible as $T \rightarrow \infty$, so we ignore it. 

We now present an upper bound on the error rate for communities (analogous to Theorem \ref{errorN}) estimated from the unweighted adjacency matrix $A$. 
For a pair of nodes $(i,j)$ such that $c_i =a$ and $c_j=b$, we have
$E[A_{ij}] = E[1\{N_{ij}>0\}] = P(N_{ij}>0)= 1- e^{( -\mu_{ab}T)}$.
Now $A$ is a $n \times n$ symmetric matrix whose elements $A_{ij}$ are independent Bernoulli random variables with mean $E[A_{ij}]$. Let $\Delta =\max\{ n \max_{i,j} E[A_{ij}],c_0 \log n\} $ for some constant $c_0$, and note that $n \max_{i,j} E[A_{ij}] =  n \max (1- \exp( -\mu_{ab}T)) = n(1- \exp( -\mu_{\max}T))$, where $\mu_{\max} = \max_{a,b} \mu_{ab}$.
Further, let $\lambda_{\min}(E[A])$ denote the minimum in absolute value non-zero eigenvalue of the matrix $E[A]$ and $|a|_{\max}$ denote the size of the largest community. Then we have the following upper bound on the error rate of spectral clustering performed on $A$.

\begin{thm}
Let $Y \sim \text{CHIP}(C,n,k,\mu,\alpha,\beta)$. 
Then, with probability at least $1-n^{-r}$, the misclustering error rate for spectral clustering on the binary adjacency matrix $A$ at time $T$ is
\begin{equation*}
r \leq 64(2+\epsilon)\frac{|a|_{\max}kc\Delta}{n(\lambda_{\min}(E[A]))^2},
\end{equation*}
where $\epsilon>0$ is a constant and $c>0$ is a constant dependent on $c_0$ and $r$.
\label{error}
\end{thm}

\subsubsection{Simplified Special Case}
\label{sec:SpecialCase}
The upper bounds on the error rates in Theorems \ref{errorN} and \ref{error} are not very informative in terms of their dependencies on key model parameters. 
In Section \ref{sec:analysis_comm}, we considered a simplified special case that allowed us to simplify the constants in Theorem \ref{errorN}, resulting in Theorem \ref{cor:misclusWeighted}, which bounds the misclustering error rate on the weighted adjacency matrix $N$. 
Similarly, we have the following result for spectral clustering using the unweighted adjacency matrix $A$.

\begin{thm}
\label{binary}
Let $Y \sim \text{CHIP}(C,n,k,\mu_1,\alpha_1,\beta_1, \mu_2, \alpha_2, \beta_2)$. The misclustering error rate for spectral clustering on the binary adjacency matrix $A$ at time $T$ is
\begin{equation}
\label{eq:binary}
r \lesssim  \frac{k^2}{n}\frac{1- \exp( -\mu_{1}T)}{(\exp( -\mu_{2}T)-\exp( -\mu_{1}T))^2}.
\end{equation}
If further we assume $\mu_1 \asymp \mu_2 \asymp o(1/T)$, such that $\mu_1 T =o(1)$ and $\mu_2T=o(1)$, then we have
\begin{equation}
\label{eq:binary_taylor}
r \lesssim \frac{nT \mu_1}{(n/k)^2(\mu_1-\mu_2)^2T^2} \asymp  \frac{k^2}{nT}\frac{\mu_1 }{(\mu_1-\mu_2)^2},
\end{equation}
whereas, for $\mu_1 \asymp \mu_2 \asymp \omega(1/T) $, such that $\mu_1 T \rightarrow \infty$ and $\mu_2 T \rightarrow \infty $, then the upper bound for the misclustering rate in Theorem \ref{error} goes to 1.
\end{thm}

We note that if the parameters are kept constant as a function of $T$, then $\mu_1 T \rightarrow \infty$ and $\mu_2 T \rightarrow \infty $. Consequently, without $k$ and $n$ changing the upper bound on the error rate for the unweighted adjacency matrix in Theorem \ref{error} explodes and becomes close to $1$, making the upper bound guarantee useless. While this result might be a drawback of the upper bound result itself, we note that unbounded error makes sense because in this regime almost all node pairs have at least one communication with high probability. Hence the unweighted adjacency matrix has a $1$ in almost all entries, and the community structure cannot be detected from this matrix. In that case, we predict that using the weighted adjacency matrix $N$ can lead to smaller error. 
Theorem \ref{cor:misclusWeighted} provides the corresponding upper bound for error rate for $N$.

The density of the aggregate adjacency matrix is governed by the parameters  of the CHIP model. Hence, to further characterize the dependence of the $\mu$ parameters on the number of nodes $n$ and time $T$ in the network, assume $\mu_1=c_1\frac{1}{f(n)g(T)}$ and $\mu_2=c_2\frac{1}{f(n)g(T)}$, where $c_1$ and $c_2$ are constants that do not depend on $n$ or $T$. Also assume $1-\alpha_1/\beta_1$ and $1-\alpha_2/\beta_2$ do not depend on $n$ and $T$. Then the upper bound on the error rate becomes $
r \lesssim \frac{k^2f(n)g(T)}{nT(c_1-c_2)^2}$. 
Now we note that consistent community detection is possible as long as $k=o\Big(\frac{\sqrt{nT}|c_1-c_2|}{f(n)g(T)}\Big)$. For example, if we set $g(T) \asymp T $ and $f(n)=\frac{n}{\log n}$, such that $\mu_1 \asymp \mu_2 \asymp \frac{\log n}{nT}$, then the expected number of events between a node pair is $O(\frac{\log n}{n})$. In that case, $r(T) \lesssim \frac{k^2}{\log n (c_1-c_2)^2}$, and consistent community detection is possible as long as $k=o(\sqrt{\log n}|c_1-c_2|)$.

A second example is where we set $g(T) \asymp 1$ and $f(n)=\frac{n}{\log n}$, such that $\mu_1 \asymp \mu_2 \asymp \frac{\log n}{n}$. The expected number of events between a vertex pair is then $O(\frac{T \log n}{n})$ and total expected number of events in the whole network is $O(n T\log n)$. In that case $r \lesssim \frac{k^2}{T \log n (c_1-c_2)^2}$, and consistent community detection is possible as long as $k=o(\sqrt{T \log n}|c_1-c_2|)$.

\subsubsection{Comparison Between Weighted and Unweighted Adjacency Matrices}
\label{sec:adjComparison}
We compare the bounds on the error rates in unweighted and weighted adjacency matrices in Theorems  \ref{binary} and \ref{cor:misclusWeighted} in the sparse regime where $\mu_1 T$ and $\mu_2 T$ are small such that we can apply the Taylor series approximation. From Theorem \ref{binary}, we have the error rate using the unweighted adjacency matrix is upper bounded by $\frac{k^2}{nT}\frac{\mu_1 }{(\mu_1-\mu_2)^2}$, while the error rate for the weighted adjacency matrix is upper bounded by 
\[\frac{k^2}{nT}\frac{\frac{\mu_1}{(1-m_1)^3 } + \frac{\mu_2}{(1-m_2)^3}}{ (\frac{\mu_1}{(1-m_1) }-\frac{\mu_2}{(1-m_2) })^2} .
\]
We can make the following comparison comments on the basis of these upper bounds. 
\begin{enumerate}
    \item If $m_1=m_2=m$ such that the community structure is expressed only through $\mu_1$ and $\mu_2$, then the error for the weighted adjacency matrix is bounded by  $\frac{k^2}{nT}\frac{\mu_1 + \mu_2 }{(\mu_1-\mu_2)^2} \frac{1}{1-m} $. This upper bound is higher than the corresponding upper bound for spectral clustering in unweighted adjacency matrix indicating a possible advantage of using the unweighted adjacency matrix.
    \item If $\mu_1= \mu_2$ such that the community structure is expressed purely through $\alpha, \beta$, then the error for the unweighted case is unbounded. However, the error for the weighted case can still be bounded, indicating a possible advantage of the weighted adjacency matrix.

\end{enumerate}

\subsubsection{Proofs}
\label{sec:proof_comm}
We begin with the proofs of Theorems \ref{errorN} and \ref{error} for spectral clustering applied to the weighted and unweighted adjacency matrices, respectively, in the general CHIP model. 
We then present the proofs of Theorems \ref{cor:misclusWeighted} and \ref{binary} for the simplified special case.

\subsubsection*{Proof of Theorem \ref{errorN}}
\begin{proof}
We start with the following result. 
\begin{lem}
Let $Y \sim \text{CHIP}(C,n,k,\mu,\alpha,\beta)$.  Let  $N$ denote the weighted adjacency matrix obtained by aggregating $Y$ at time $T \rightarrow \infty$. Then, with probability at least $1-1/n$, we  have
\begin{equation}
\label{eq:bandeira_bound}
\|N-E[N]\|_2 \leq (1+\epsilon)\left\{2s + \frac{6}{\log (1+\epsilon)}s_1\sqrt{\log n}\right\} + 2s_1\sqrt{\log n},
\end{equation}
where $0<\epsilon<1/2$ is a constant, and the terms $s$ and $s_1$ are as defined in \eqref{eq:const_s} and \eqref{eq:const_s1}, respectively.
\label{boundN}
\end{lem}
We present the proof of this lemma following the proof of this theorem.

Since $E[N]$ can also be written in the form of a stochastic block model as $E[N] = C\nu C^T$, we can use the same arguments as in the proof of the previous result. Using the Davis-Kahan Theorem~\citep{dk70,stewart},  we have the following bound:
\begin{align}
r &\leq \frac{1}{n}|a|_{\max}8 (2+\epsilon_1)\|\hat{U}-C(C^TC)^{-1/2}\mathcal{O}\|_F^2 \nonumber \\
&\leq 64(2+\epsilon_1)\frac{|a|_{\max}k\|N-E[N]\|_2^2}{n(\lambda_{\min}(E[N])^2} ,
\label{misclus}
\end{align}

Combining \eqref{eq:bandeira_bound} and \eqref{misclus}, we arrive at the desired result.

\end{proof}

\subsubsection*{Proof of Lemma \ref{boundN}}
\begin{proof}
We note that $N_{ij}$ is asymptotically normal (Theorem 4 of \cite{hawkes1974cluster}) as $T \rightarrow \infty$, i.e.
\[
N_{ij} |(C_{ia}=1, C_{jb}=1) \sim \mathcal{N} ( \nu_{ab}, \sigma^2_{ab}).
\]
Then $(N-E[N])$ is a $n\times n$ symmetric matrix with elements $(N-E[N])_{ij}=g_{ij}\sigma_{ij}$, where $g_{ij}; i\geq j$ are i.i.d $\mathcal{N}(0,1)$ and $\sigma_{ij}$ is the standard deviation of $N_{ij}$ given before. 

We will use Corollary 3.9 in \cite{bandeira2016sharp}. In the notation of \cite{bandeira2016sharp}, we set $\sigma =s $, $\sigma^{*}=s_1$ and let $t= 2s_1\sqrt{\log n} $. Then for any $0<\epsilon<1/2$, we have
\begin{equation*}
P\left(\|N-E[N]\|_2 \geq (1+\epsilon)\{2s + \frac{6}{\log (1+\epsilon)}s_1\sqrt{\log n}\} + 2s_1\sqrt{\log n}\right) \\ 
\leq  \exp(-\log n).
\end{equation*}
Since $\exp(-\log n) = 1/n$, one can then take the probability of the complement, which completes the proof.
\end{proof}

\subsubsection*{Proof of Theorem \ref{error}}
\begin{proof}

We note that the matrix $A$ is an adjacency matrix with independent entries. Further $n \max_{ij}  E[A_{ij}] \leq \Delta$ and $\Delta \geq c_0 \log n$ by definition. Then by Theorem 5.2 of  \cite{lei2015consistency}, we have with probability at least $1-n^{-r}$,
\begin{equation}
\label{eq:bound_Lei}
\|A-E[A]\|_2 \leq c \sqrt{\Delta},
\end{equation}
where $c$ is a constant dependent on $c_0$ and $r$. 

Since $E[A]$ can be written in the form of a stochastic block model as $E[A] = C(1-\exp(\mu T)) C^T$, we can use known results in the SBM literature. Let $\hat{U}_{n \times k}$ denote the $n \times k$ matrix whose columns are the top $k$ eigenvectors of the matrix $A$. By Lemma 3.1 of \citet{rcy11}, the matrix of eigenvectors corresponding to the largest $k$ non-zero eigenvalues of the matrix $E[A]$ is $C(C^TC)^{-1/2}\mathcal{O}$ for some $k \times k$ orthogonal matrix $\mathcal{O}$. Then we have the following relationship for the difference between matrices of population eigenvectors (those of $E[A]$) and sample eigenvectors (those of $A$) and the misclustering error rate of community detection by applying $(1+\epsilon)$ approximate $k$-means algorithm to those matrices \citep{pensky2019spectral}:
\begin{equation}
r \leq \frac{1}{n}|a|_{\max}8 (2+\epsilon)\|\hat{U}-C(C^TC)^{-1/2}\mathcal{O}\|_F^2.
\end{equation}
Next we use the Davis-Kahan Theorem~\citep{dk70,stewart} that relates perturbation of matrices to perturbation of eigenspaces of those matrices. Then we have the following bound on the misclustering rate (also see Lemma 5.1 of \cite{lei2015consistency}):
\begin{equation}
r \leq  64(2+\epsilon)\frac{|a|_{\max}k\|A-E[A]\|_2^2}{n(\lambda_{\min}(E[A])^2}.
\label{misclus_A}
\end{equation}
 
Combining \eqref{eq:bound_Lei} and \eqref{misclus_A}, we arrive at the desired result.

\end{proof}

Next, we present the proofs of the theorems for the simplified special case with $k$ equivalent communities. 

\subsubsection*{Proof of Theorem \ref{cor:misclusWeighted}}

\begin{proof}
Under the simplified model we have
\[
E[N] = C\left((\nu_1-\nu_2)TI_k + \nu_2T \vec{1}_k \vec{1}_k^T\right)C^T.
\]
As before all communities have the same number of nodes, i.e., $|a| = \frac{n}{k}$ for all $a$, and $|a|_{\max} =\frac{n}{k}$. Then by \citet{rcy11}, $\vec{1}_k$ is an eigenvector corresponding to the eigenvalue $\frac{n}{k}(\nu_1-\nu_2)T+n\nu_2 T,$ and the remaining non-zero eigenvalues are of the form $\frac{n}{k}(\nu_1-\nu_2)T$.  Since $n\nu_2>0$, the smallest non-zero eigenvalue 
\begin{equation*}
\lambda_{\min}(E[N]) = \frac{n}{k}(\nu_1-\nu_2)T.
\end{equation*}

The upper bound from Theorem \ref{errorN} can also be simplified further under this model. We have
\[
s= \sqrt{T} \sqrt{\frac{n}{k}\sigma^2_1 + \frac{(k-1)n}{k} \sigma^2_2} \asymp \sqrt{\frac{nT}{k}}\sqrt{\sigma_1^2 + (k-1)\sigma_2^2} \asymp \sqrt{nT}\sigma_1,
\]
and
\[
s_1 = \sqrt{T} \sigma_1,
\]
and consequently,
\begin{align*}
(1+\epsilon)\left(2s + \frac{6}{\log (1+\epsilon)}s_1\sqrt{\log n}\right) + 2s_1\sqrt{\log n} &\asymp \sqrt{T} \sigma_1 \left(\sqrt{n} + \sqrt{\log n}\right) \\
&\lesssim  \sqrt{T} \sigma_1 \sqrt{n}.
\end{align*}
Substituting these quantities into Theorem \ref{errorN} completes the proof.

\end{proof}

\subsubsection*{Proof of Theorem \ref{binary}}

\begin{proof}
Under the simplified model all communities have the same number of nodes, i.e., $|a| = \frac{n}{k}$ for all $a$, and consequently $|a|_{\max} =\frac{n}{k}$. Further, we can write
\[
E[A] = C\left((\exp( -\mu_{2}T)-\exp( -\mu_{1}T))I_k + (1- \exp( -\mu_{2}T) \vec{1}_k \vec{1}_k^T\right)C^T,
\]
where  $I_k$ is the $k$-dimensional identity matrix, and $\vec{1}_k$ is the $k$-dimensional vector of all $1$'s. Then by \citet{rcy11}, $\vec{1}_k$ is an eigenvector corresponding to the eigenvalue $\frac{n}{k}(\exp( -\mu_{2}T)-\exp( -\mu_{1}T))+n(1- \exp( -\mu_{2}T))$, and the remaining non-zero eigenvalues are of the form $\frac{n}{k}(\exp( -\mu_{2}T)-\exp( -\mu_{1}T))$.  Since $n(1- \exp( -\mu_{2}T))>0$, the smallest in absolute value non-zero eigenvalue of $E[A]$ is then, 
\begin{equation*}
\lambda_{\min}(E[A]) = \frac{n}{k}(\exp( -\mu_{2}T)-\exp( -\mu_{1}T)).
\end{equation*}
Also, under this setting, the numerator in the upper bound from Theorem \ref{error} becomes 
\begin{equation*}
\Delta = n (1- \exp( -\mu_{1}T)). 
\end{equation*}
Substituting these quantities into Theorem \ref{error}, we arrive at \eqref{eq:binary}, the first statement of the theorem.

If we further assume that $\mu T $ is small then we can make some further simplifications using the Taylor series expansion of $\exp(-x)$ near $x=0$. 
In this case,
\begin{equation*}
\lambda_{\min} \asymp \frac{n}{k} ( \mu_1 - \mu_2)T, 
\end{equation*}
and
\begin{equation*}
\Delta \asymp n\mu_1T.
\end{equation*}
Substituting these quantities into Theorem \ref{error}, we arrive at \eqref{eq:binary_taylor}, the second statement of the theorem, which completes the proof.
\end{proof}

\subsection{Estimated Hawkes Process Parameters}
\label{sec:proof_param}

\subsubsection{Confidence Intervals}
\label{sec:HawkesParamConfInterval}
We derive confidence intervals for $m$ using Theorem \ref{hawkes_thm} and the following result readily obtained using the Law of Large numbers: $\bar{N}_{ab} \overset{p}{\rightarrow} \mu_{ab}$. A $(1-\theta)*100 \%$ Bonferroni-corrected (due to multiple comparisons) simultaneous confidence interval for all $k^2$ parameters $m_{ab}$ is
\begin{equation}
\label{eq:m_conf_int}
\hat{m}_{ab}  \pm z_{(1-\frac{\theta}{2k^2})}\sqrt{\frac{1}{4n_{ab}\bar{N}_{ab}}}.
\end{equation}
The confidence intervals on $m_{ab}$ are particularly appealing to detect the ``burstiness" of the network dynamics by testing the hypothesis $m_{ab} > 0$ for a block pair $(a,b)$.

For the $\mu$ parameters, we are more interested in confidence intervals for pairwise differences between block pairs to identify whether the block pairs differ in their baseline event rates. Therefore, we build the following pairwise confidence intervals for all $2k(k-1)$ pairwise differences: 
\begin{equation}
\label{eq:mu_diff_conf_int}
(\hat{\mu}_{ab} - \hat{\mu}_{ac})  \pm z_{(1-\frac{\theta}{4(k-1)k})}\frac{1}{T}\sqrt{\frac{9}{4}\left(\frac{\bar{N}_{ab}}{n_{ab}} + \frac{\bar{N}_{ac}}{n_{ac}}\right)}.
\end{equation}

Note even though the random variables $\hat{m}_{ab}$ and $\hat{\mu}_{ab}$ are dependent across block pairs due to the spectral clustering step, the Bonferroni correction is still going to give a conservative (wide) interval with a simultaneous confidence coverage at least $1-\theta$. 

\subsubsection{Proofs} \label{sec:estimatedHawkesParamProofs}

\subsubsection*{Proof of Theorem \ref{hawkes_thm}}
\begin{proof}
First, using the Central Limit Theorem and Law of Large Numbers, we have
\[
\bar{N}_{ab}  \overset{d}{\rightarrow} \mathcal{N}\left(\nu_{ab},\frac{\sigma^2_{ab}}{n_{ab}}\right) \text{ and } S^2_{ab} \overset{p}{\rightarrow} \sigma^2_{ab}, \quad \quad \text{ as } n_{ab}\rightarrow \infty.
\]
Then by Slutsky's theorem \citep{lehmann2004elements} we have,
\[
\frac{\bar{N}_{ab}}{S^2_{ab}} \overset{d}{\rightarrow} \mathcal{N}\left(\frac{\nu_{ab}}{\sigma^2_{ab}},\frac{1}{\sigma^2_{ab}n_{ab}}\right) \Leftrightarrow \sqrt{n_{ab}}\left(\frac{\bar{N}_{ab}}{S^2_{ab}}-\frac{\nu_{ab}}{\sigma^2_{ab}}\right) \overset{d}{\rightarrow} \mathcal{N}\left(0,\frac{1}{\sigma^2_{ab}}\right).
\]
Finally, we will apply the delta method (See Theorem 2.5.2 of \cite{lehmann2004elements}) on the random variable $X = \frac{\bar{N}_{ab}}{S^2_{ab}} $ with the function $g(x)= 1-\sqrt{x}$. Note that $g'(x)=\frac{1}{2\sqrt{x}}$. Then we can compute $g'\left(\frac{\nu_{ab}}{\sigma^2_{ab}}\right) = \frac{\sigma_{ab}}{2\sqrt{\nu_{ab}}}$.  Then we have
\[
\sqrt{n_{ab}} \left(\hat{m}_{ab} - \left(1-\sqrt{\frac{\nu_{ab}}{\sigma^2_{ab}}}\right)\right) \overset{d}{\rightarrow} \mathcal{N} \left(0,\frac{1}{4\nu_{ab}}\right).
\]

Next we derive the asymptotic distribution for $\hat{\mu}_{ab}$. We first apply the delta method to the random variable $\bar{N}_{ab}$ with the function $g(x)=x^{3/2}$. Clearly, $g'(x) = \frac{3}{2}\sqrt{x}$, such that $g'(\nu_{ab}) = \frac{3}{2}\sqrt{\nu_{ab}}$. Then we have
\[
\sqrt{n_{ab}}((\bar{N}_{ab})^{3/2} - (\nu_{ab})^{3/2}) \overset{d}{\rightarrow} \mathcal{N} \left(0,\frac{9}{4}\nu_{ab}\sigma^2_{ab}\right).
\]
Applying Slutsky's theorem, we then have
\[
\sqrt{n_{ab}}\left(\frac{(\bar{N}_{ab})^{3/2}}{S_{ab}} - \frac{(\nu_{ab})^{3/2})}{\sigma_{ab}}\right) \overset{d}{\rightarrow} \mathcal{N} \left(0,\frac{9}{4}\nu_{ab}\right).
\]
\end{proof}

\subsubsection*{Proof of Theorem \ref{endtoend}}
Let $\bar{C}$ and $\hat{C}$ denote the true and estimated community assignment matrices respectively. Define $\bar{H}= \bar{C} (\bar{C}^T\bar{C})^{-1/2}$ and 
$\hat{H}= \hat{C} (\hat{C}^T\hat{C})^{-1/2}$, such that $\bar{H}^T\bar{H}=\hat{H}^T\hat{H} = I$.

We have 
\[
E[N]=\bar{C} \nu \bar{C}^T
\]
Then 
\[
(\bar{C}^T\bar{C})^{1/2}\nu (\bar{C}^T\bar{C})^{1/2} = (\bar{C}^T\bar{C})^{-1/2} \bar{C}^T E[N] \bar{C} (\bar{C}^T\bar{C})^{-1/2} =\bar{H}^T E[N]\bar{H}.
\]
Instead, the estimate for $\nu$ we get using estimated community assignment matrix $\hat{C}$ applied to $N$ is
\[
(\hat{C}^T\hat{C})^{1/2}\hat{\nu} (\hat{C}^T\hat{C})^{1/2} = \hat{H}^T N \hat{H}
\]

Note that $(\hat{C}^T\hat{C})$ and $(\bar{C}^T\bar{C})$ are $k \times k$ diagonal matrices whose $q$th diagonal element represents the number of vertices that are part of the $q$th community. Next we make a key assumption---the sizes of the communities from the estimated community partition are similar to the true community sizes. In particular, we assume that the size of each of the $k$ communities in the true and estimated partition is $O(\frac{n}{k})$.   Therefore, the difference
\[
(\hat{C}^T\hat{C})^{1/2}\hat{\nu} (\hat{C}^T\hat{C})^{1/2} - (\bar{C}^T\bar{C})^{1/2}\nu (\bar{C}^T\bar{C})^{1/2} \asymp \frac{n}{k} (\hat{\nu}-\bar{\nu}).
\]

Now we have
\begin{align*}
 \frac{n}{k} (\hat{\nu}- \nu) &= \hat{H}^TN\hat{H} - \bar{H}^TE[N]\bar{H} \\
& = \hat{H}^TN\hat{H} - \hat{H}^TE[N]\hat{H} + \hat{H}^TE[N]\hat{H} - \bar{H}^TE[N]\bar{H} \\
& =  \hat{H}^T(N-E[N])\hat{H} + \{ \hat{H}^TE[N](\hat{H} - \bar{H}) + (\hat{H} - \bar{H})^T E[N] \bar{H} \}
\end{align*}
We also note that 
\[
\|\bar{H}\|_2 \leq \sqrt{\lambda_{\max}(\bar{H}^T\bar{H})} = 1,
\]
Note by assumption, $\sqrt{n_{ab}} \asymp \frac{n}{k}$.
Now, 
\begin{align*}
   \sum_{ab} n_{ab}(\hat{\nu} -\nu)_{ab}^2 & \asymp \left(\frac{n}{k}\right)^2\|\hat{\nu} -\nu\|_F^2 \\
    & \leq ( \|\hat{H}^T(N-E[N])\hat{H}\|_F + 2\| \hat{H}^TE[N](\hat{H} - \bar{H})\|_F )^2\\
    & \leq 2k\left(\|N-E[N]\|_2^2 + 4\|E[N]\|_2^2 \frac{\|N-E[N]\|_2^2}{\lambda^2_{min}(N)}\right)
\end{align*}

In the notation of Theorem \ref{cor:misclusWeighted}, 
\[
\lambda_{\min}(N) = \frac{n}{k}(\nu_1 -\nu_2)T.
\]
Also, using the upper bound in terms of expectation (instead of the in probability upper bound) from Theorem \ref{cor:misclusWeighted} we have 
\[
E[\|N-E[N]\|_2] \lesssim \sqrt{nT}\sigma_1, \quad \|E[N]\|_2 \lesssim n\nu_1 T.
\]
Therefore,
\[
E\left[\sum_{ab} n_{ab}(\hat{\nu} -\nu)_{ab}^2\right] \lesssim knT\sigma_1^2 + k \frac{n^2\nu_1^2T^2nT\sigma_1^2k^2}{n^2(\nu_1 -\nu_2)^2T^2} \lesssim knT\sigma_1^2 + \frac{k^{3}nT\sigma_1^2\nu_1^2}{(\nu_1-\nu_2)^2}.
\]
And consequently the sum of the weighted mean squared errors is,
\[
\sum_{ab}n_{ab}E[(\hat{\nu} -\nu)_{ab}^2] \lesssim knT \max\left\{\sigma_1^2, \frac{k^2\sigma_1^2\nu_1^2}{(\nu_1-\nu_2)^2} \right\}
\]
Noting that $\sum_{ab}n_{ab} \asymp n^2$, the average MSE of estimating $\nu_{ab}$ is then asymptotically
\[
\frac{kT}{n} \max\left\{\sigma_1^2, \frac{k^2\sigma_1^2\nu_1^2}{(\nu_1-\nu_2)^2}\right\} \text{   or   } \frac{T}{\sqrt{n_{ab}}} \max\left\{\sigma_1^2, \frac{k^2\sigma_1^2\nu_1^2}{(\nu_1-\nu_2)^2}\right\} 
\]

For comparison, the weighted sum of MSEs in estimating $\nu_{ab}$, using the estimator $\bar{N}_{ab}$ when the community structure is known (from Theorem \ref{hawkes_thm}) is 
\[
\sum_{ab} n_{ab} E[(\bar{N}_{ab} - \nu_{ab})^2]  = \sum_{ab}\sigma_{ab}^2  = k^2T\sigma_1^2,
\]
and average MSE is asymptotically
\[\frac{k^2T\sigma_1^2}{n^2} \text{ or } \frac{T\sigma_1^2}{n_{ab}}.
\]

\section{Additional Experiments} \label{sec:additionalExperiments}

We present two additional simulation experiments to analyze the effects of various parameters of the CHIP model on the accuracy of spectral clustering and to compare spectral clustering using weighted and unweighted adjacency matrices in detecting the ground truth community structure in simulated networks. 
We then present additional details and analyses for our real network dataset experiments. 

\subsection{Simulation Experiments}

\subsubsection{Community Detection with Varying \texorpdfstring{$n$}{n}} \label{sec:sc_weighted_vs_unweighted}
\begin{figure}[t]
    \centering
    \subfigure[Only $\mu$ is informative ($\mu_1 \neq \mu_2$ and $\alpha_1 = \alpha_2$)]{
    \includegraphics[width=2in]{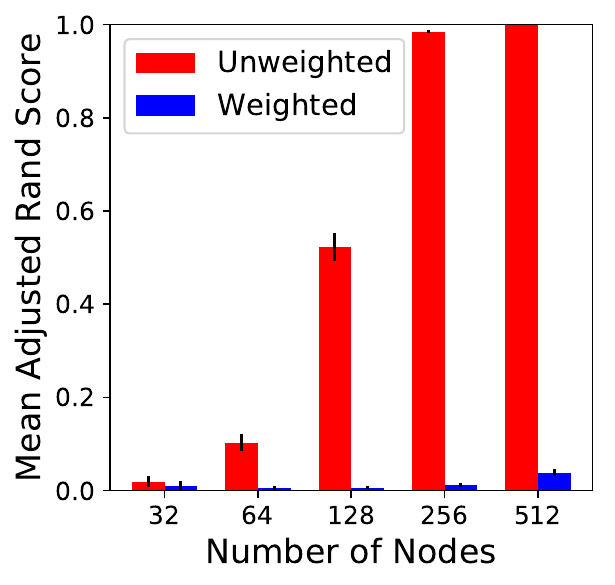}
    \label{fig:commDetectionMu}
    }
    \quad
    \subfigure[Only $\alpha$ is informative ($\mu_1 = \mu_2$ and $\alpha_1 \neq \alpha_2$)]{
    \includegraphics[width=2in]{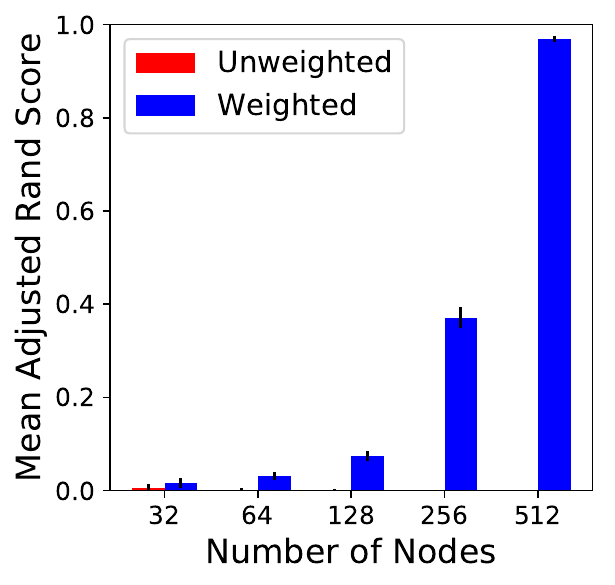}
    \label{fig:commDetectionAlpha}
    }
    \caption[Mean adjusted Rand score of spectral clustering on weighted and unweighted
adjacency matrices over 100 simulated networks]
{Mean adjusted Rand scores of spectral clustering on weighted and unweighted
adjacency matrices over 100 simulated networks ($\pm$ 2 standard errors). $\beta_1 = \beta_2$ in both cases. Both sets of results agree with upper bounds from Section \ref{sec:adjComparison}.}
\end{figure}

We simulate networks from the simplified CHIP model described in Section \ref{sec:SpecialCase} with $k=4$ communities, duration $T=400$, and a growing number of nodes $n$. 
We estimate community assignments of nodes using both the weighted adjacency (count) matrix $N$ and unweighted adjacency matrix $A$.

First, we choose parameters $\mu_1 = 0.002$, $\mu_2 = 0.001$, $\alpha_1 = \alpha_2 = 7$, and $\beta_1 = \beta_2 = 8$ so that only $\mu$ is informative. 
The upper bound on the misclustering error rate using $N$ is worse by a factor of $(1-m)^{-1} = 8$ compared to using $A$ as discussed in Section~\ref{sec:adjComparison}.
The adjusted Rand scores for spectral clustering on both $A$ and $N$ over $100$ simulated networks for varying $n$ are shown in Figure~\ref{fig:commDetectionMu}. 
The accuracy on $A$ approaches $1$ for growing $n$, as expected. 
The accuracy on $N$ is significantly worse, as predicted by the comparison of the respective upper bounds on the misclustering error rates, and no better than a random community assignment until $n=512$ nodes. 

Next, we choose parameters $\mu_1 = \mu_2 = 0.001$, $\alpha_1 = 0.006$, $\alpha_2 = 0.001$, and $\beta_1 = \beta_2 = 0.008$. 
so that only $\alpha$ is informative. 
The error for $A$ is unbounded, while the error for $N$ still follows the upper bound in Corollary \ref{cor:misclusWeighted}. As shown in Figure \ref{fig:commDetectionAlpha}, the accuracy on $N$ approaches 1 as $n$ increases, while the accuracy on $A$ is no better than random even for growing $n$, as expected.

\subsubsection{Effects of Diagonal and Off-diagonal \texorpdfstring{$\mu$'s}{Mu's} on Community Detection}
\label{increasemu}
In Section \ref{sec:sc_weighted_vs_unweighted} we observed that community detection will be easier if $\mu$ is informative ($\mu_1 \neq \mu_2$). In this experiment, we will explore two different ways of encoding community information into simulated networks by 
\begin{enumerate}
  \item Scaling up both $\mu_1$ and $\mu_2$, while keeping a fixed $\mu_1:\mu_2$ ratio.
  \item Only scaling up $\mu_1$, allowing for $\mu_1:\mu_2$ ratio to increase.
\end{enumerate}

Both settings share the same base parameters of $\mu_1 = 0.075$ and $\mu_2 = 0.065$, with $k=4$ communities and $n=128$ nodes, a duration of $T=50$, where $\alpha_1 = \alpha_2 = 0.05$ and $\beta_1 = \beta_2 = 0.08$. These parameters are chosen to create a base network that is nearly impossible for spectral clustering to accurately detect communities. The objective is similar to that of Section \ref{sec:sc_weighted_vs_unweighted}, where in both settings we perform community detection using spectral clustering on the weighted adjacency of simulated networks, while increasing $\mu_1$ and $\mu_2$ or their ratio. Lastly, we average over the adjusted Rand score of 100 simulations.

\begin{figure}[t]
    \centering
    \subfigure[Scaling both $\mu$'s up, with a fixed $\mu_1:\mu_2$ ratio]{
    \includegraphics[width=3in]{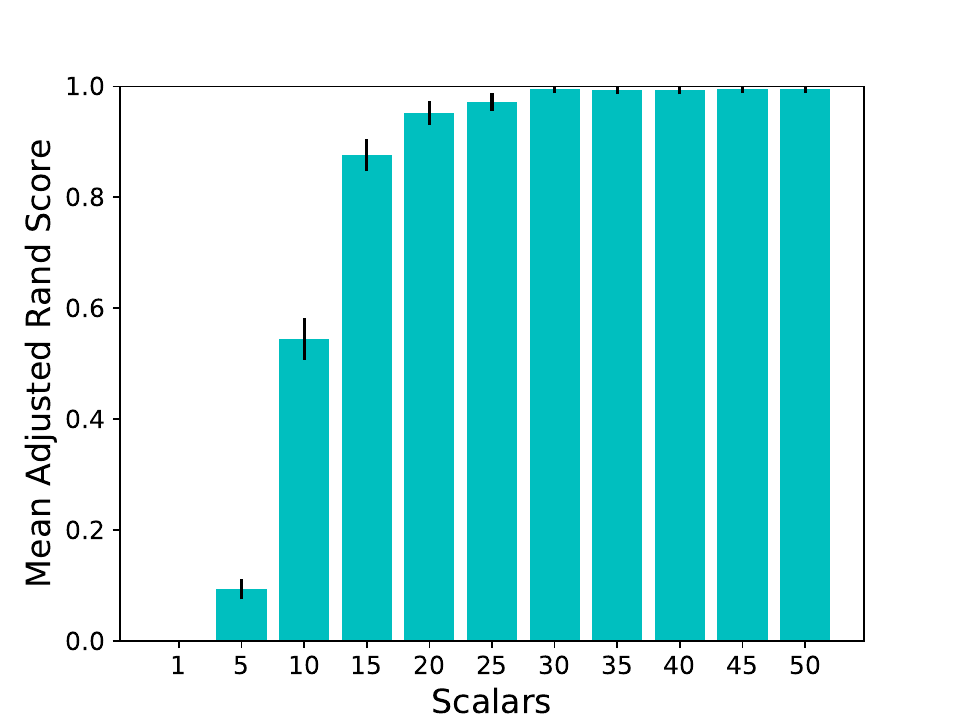}
    \label{fig:commDetectionFixedMuRatio}
    }
    \subfigure[Scaling $\mu_1$ up only, keeping $\mu_2$ fixed]{
    \includegraphics[width=3in]{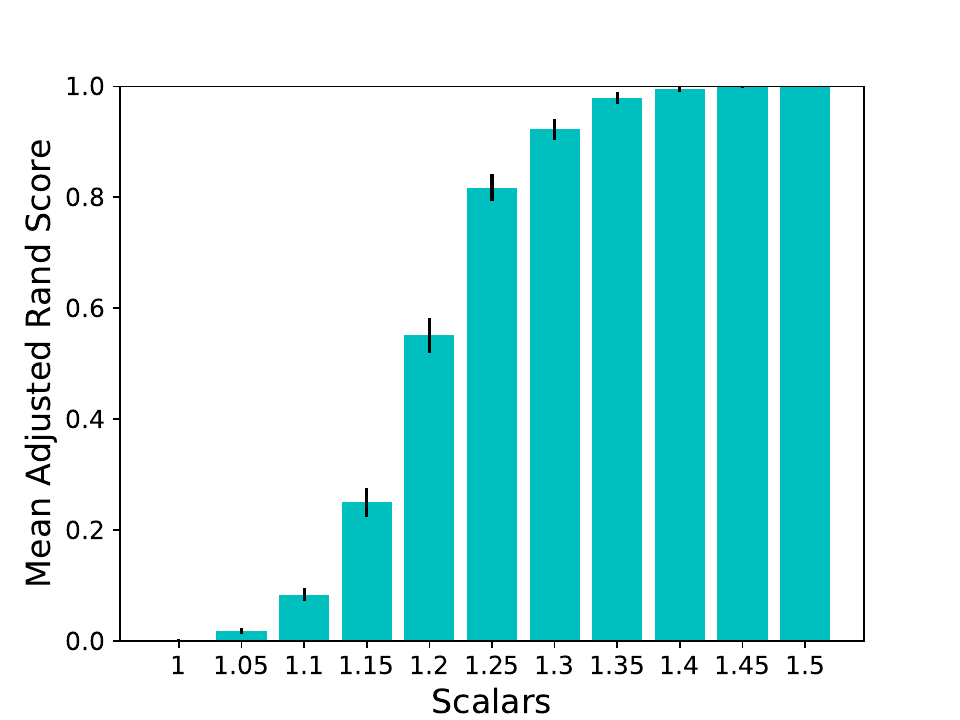}
    \label{fig:commDetectionIncreaseMu1}
    }
    
    \caption[Adjusted Rand score of spectral clustering on weighted
adjacency matrix when scaling parameters]
{Adjusted Rand score of spectral clustering on weighted
adjacency matrix, averaged over 100 simulated networks ($\pm$ 2 standard errors), while multiplying $\mu_1$ and $\mu_2$ or their ratio by scalars. \subref{fig:commDetectionFixedMuRatio} Scaling up both $\mu_1$ and $\mu_2$, keeping their ratio fixed. \subref{fig:commDetectionIncreaseMu1} Only scaling up $\mu_1$, while keeping $\mu_2$ fixed.}
\label{fig:commDetectionScalingMu}
\end{figure}

\begin{figure*}[tp]
    \centering
    \includegraphics[width=\textwidth,clip=true,trim=1.1in 0in 1.1in 0in]{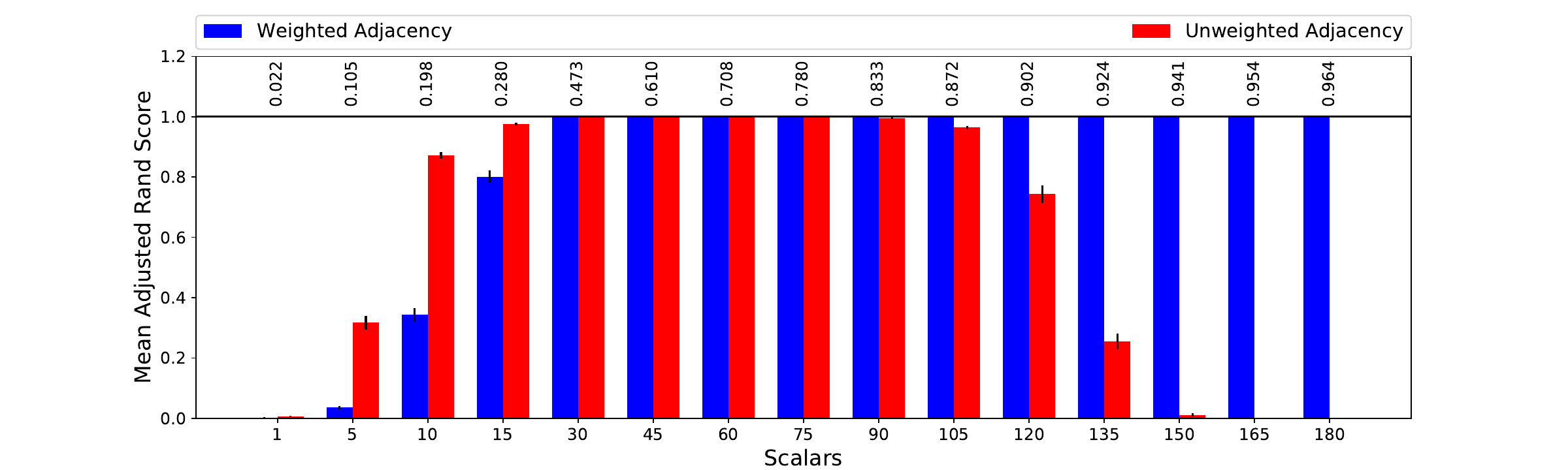}
    \caption{Adjusted Rand score of spectral clustering on weighted vs.~unweighted
adjacency matrices, averaged over 100 simulations ($\pm$ 2 standard errors), while multiplying both $\mu_1$ and $\mu_2$ by scalars. The numbers above each bar indicate the average density of simulated networks as the proportion of non-zero entries to the total number of elements in the adjacency matrix. Base model parameters are: $\mu_1 = 7.5 \times 10^{-4}$, $\mu_2 = 3.5 \times 10^{-4}$, $k=4$, $T=50$, $n=256$, $\alpha_1 = \alpha_2 = 0.05$, and $\beta_1 = \beta_2 = 0.08$.}
    
\label{fig:commDetectionIncreaseDensity}
\end{figure*}

As shown in Figure \ref{fig:commDetectionScalingMu}, community detection accuracy increases in both settings as the scalars increase; however, we find that the increase in accuracy occurs for different reasons. In the first setting, Figure \ref{fig:commDetectionFixedMuRatio}, where both $\mu$'s are scaled up with a fixed ratio, community detection becomes easier simply because the networks are becoming denser, as shown in the numbers above the bars in Figure \ref{fig:commDetectionIncreaseDensity}, and more information is available. Furthermore, although we keep the $\mu_1:\mu_2$ ratio fixed, as the scalars increase the difference between the two starts to magnify. On the other hand, as networks become denser and most node pairs start to have at least one interaction, it is only the number of interactions among node pairs that becomes informative. Therefore, spectral clustering on the weighted adjacency matrix continues to result in a high adjusted Rand score, while the adjusted Rand score of spectral clustering on the unweighted adjacency matrix decreases with increasing density, as it is illustrated in Figure \ref{fig:commDetectionIncreaseDensity}. This observation confirms the theoretical prediction made in Theorem \ref{binary}. The opposite also holds to some degree. For really sparse networks spectral clustering is more accurate on the unweighted adjacency matrix; however, in Figure \ref{fig:commDetectionIncreaseDensity} we observe that it loses its advantage as the proportion of node pairs with at least one interaction approaches 0.5 and starts impairing community detection as it passed 0.8. 

In the second setting, Figure \ref{fig:commDetectionIncreaseMu1}, by only scaling up $\mu_1$, the difference between the baseline rate of occurrence of an event between the diagonal and the off-diagonal blocks increases. This increases the signal-to-noise ratio and is a more effective way of encoding community information into a network. This can be observed by comparing the scalars of Figures \ref{fig:commDetectionFixedMuRatio} and \ref{fig:commDetectionIncreaseMu1}. Starting from the same base network, a perfect adjusted Rand score is achieved when only $\mu_1$ is scaled up by a factor of $1.4$, compared to scaling both $\mu$'s up by a factor of 30.

\subsection{Real Data}

\subsubsection{Dataset Descriptions} \label{sec:datasetDescriptions}

We consider three real network datasets consisting of timestamped relational events.  
For each dataset, we normalize the event times to the range $[0,1,\!000]$.

\begin{itemize}
    \item MIT Reality Mining \citep{eagle2009inferring}: Consists of 2,161 phone calls where the start time of each call was used as the event timestamp. This dataset has a ``core-periphery'' structure, where there is a core group for whom we have all of their communication data and a much larger group of people in the periphery who had contact with the core. 
    We consider calls between pairs of the core 70 callers and recipients. We use the last 661 phone calls as the test set\footnote{We found some inconsistencies between the actual dataset used and its description in \cite{Dubois2013}. For a fair comparison, we loaded and preprocessed this dataset using their code available on GitHub: \url{https://github.com/doobwa/blockrem/blob/master/process/reality.r}.}.

    \item Enron \citep{klimt2004enron}: Consists of 4,000 emails exchanged among 142 individuals. We use the last 1,000 emails as the test set.

    \item Facebook Wall Posts \citep{viswanath2009evolution}: Consists of a total of 876,993 wall posts from 46,952 users  from September 2004 to January 2009. 
    We consider only posts from a user to another user's wall so that there are no self-edges.
    We analyze the largest connected component of the network excluding self loops: 43,953 nodes and 852,833 events. We divide the dataset into train and test sets using a 80\%/20\% split on the number of events.
\end{itemize}

\subsubsection{Comparison with Other Models}

We find that our proposed CHIP model achieves higher test log-likelihood than the relational event model (REM) \citep{Dubois2013}, block Hawkes model (BHM) \citep{junuthula2017block}, and the spectral clustering with homogeneous Poisson process baseline on the Reality Mining and Enron datasets as shown in Table \ref{tab:realDataModelComparison}. 
CHIP and the Poisson baseline were able to scale to the Facebook network, which was two orders of magnitude larger. 
The local search procedure in the BHM does not scale to such a large network, so we provide fits using only the faster but less accurate spectral clustering procedure. 
We did not implement the REM so we compare against the reported results in \citet{Dubois2013}, which did not include the Facebook data.
We note that since all the three models assume the same Poisson process for arrival of events with different rates (which are governed by different set of parameters), the joint distribution of event times has the same form for all three models. Hence the likelihood function of the models are directly comparable. Therefore, the test log-likelihood is a reasonable metric for comparing the fits of the models to the data.

To compute the test log-likelihood for CHIP and BHM, we use the following process.
First, we use the estimation procedure explained in Section \ref{sec:EstimationProcedure} to estimate all CHIP's Hawkes process parameters using the training set (the entire dataset excluding the test set). Next, we calculate the model log-likelihood on the entire dataset and subtract the training log-likelihood from it. The result is then divided by the total number of events in the test set to evaluate the mean log-likelihood per test event, which is the metric used in \citet{Dubois2013}. 
Lastly, if a node in the test set did not appear in the training data, it was automatically assigned to the largest block.

We implemented the BHM by using spectral clustering followed by local search \citep{junuthula2017block}, which they found 
to achieve the highest adjusted Rand score in simulations compared to just spectral clustering and variational EM. 
We allowed the local search to converge to a local maximum for all values of $k$.

\subsubsection{Exploratory Analysis of Enron Network} \label{sec:enronExploratoryAnalysis}

\begin{figure}[tp]
	\centering
    \includegraphics[width=2.5in]{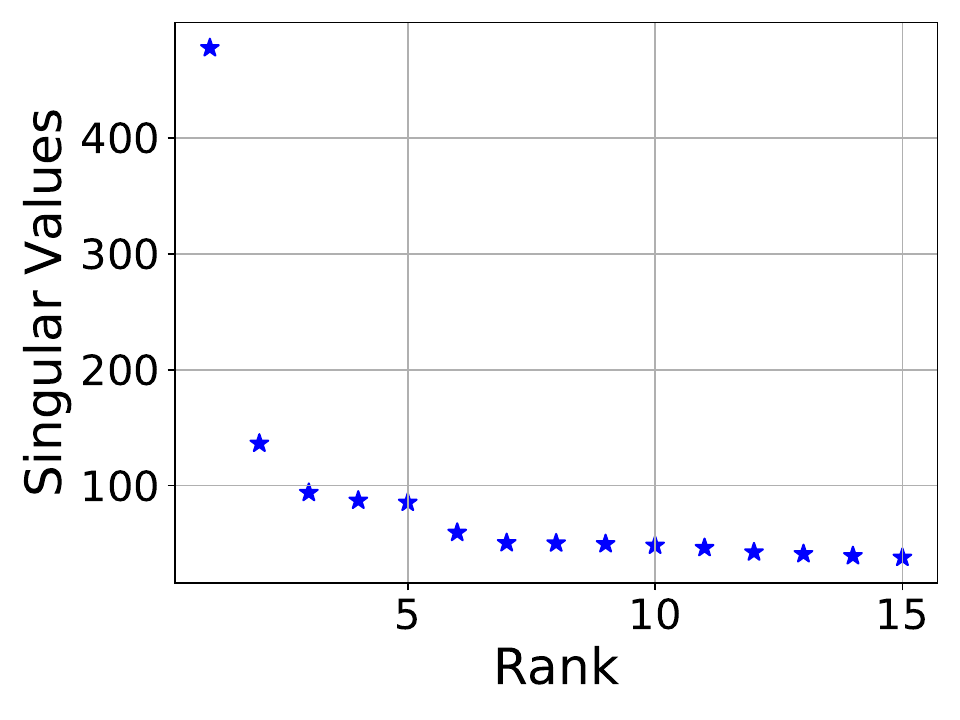}
    \caption[15 largest singular values of spectral clustering on the weighted adjacency matrix of the Enron dataset]{15 largest singular values of spectral clustering on the weighted adjacency matrix of the Enron dataset. The gap between the 2\textsuperscript{nd} and the 3\textsuperscript{rd} largest singular values led us to select $k=2$ blocks.}
    \label{fig:enronSingular}
\end{figure}

Next, we perform model-based exploratory analysis of the Enron network using CHIP. 
We find a large gap between the 2\textsuperscript{nd} and the 3\textsuperscript{rd} largest singular values of the weighted adjacency matrix as shown in Figure \ref{fig:enronSingular} so we choose a fit with $k=2$ blocks.
The number of node pairs and events in each block pair are shown in Table \ref{tab:enronNumNodesAndEvents}.

\begin{table}[tp]
     \centering
     \caption{Number of node pairs and events in each block pair of the CHIP model with $k=2$ in the Enron dataset.}
     \label{tab:enronNumNodesAndEvents}
     \begin{tabular}{r|cccc}
         \toprule
         \textbf{Block Pair (a, b)} & (1, 1) & (1, 2) & (2, 1) & (2, 2) \\ \hline
         \textbf{Node Pair Count}  & 5700   & 5016  & 5016   & 4290   \\
         \textbf{Event Count}       & 965    & 572   & 1038   & 1425   \\
         \bottomrule
     \end{tabular}
 \end{table}

\begin{figure}[tp]
    \centering
    \subfigure[$\hat{\mu}$]{
    \includegraphics[width=1.5in]{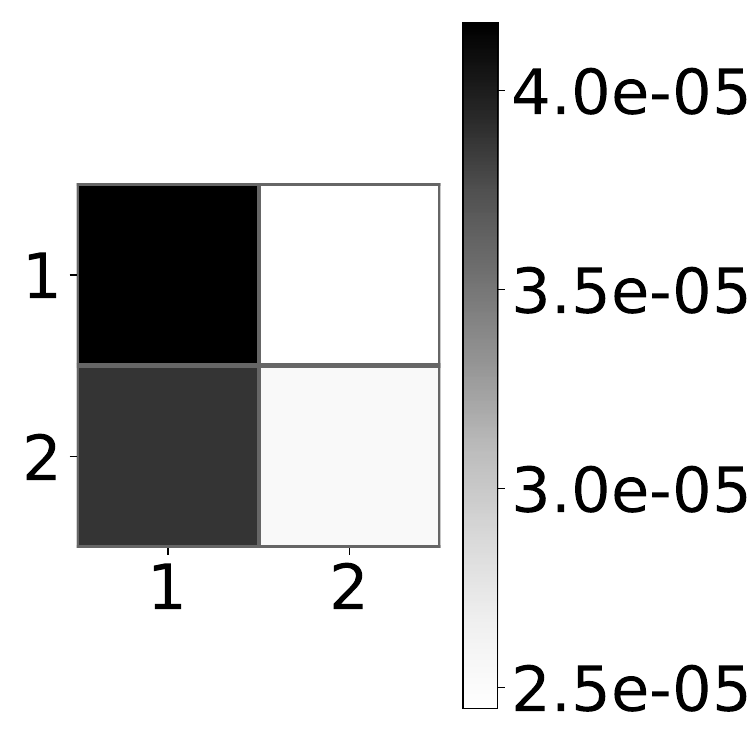}
    \label{fig:enronChpMu}
    }%
    \subfigure[$\hat{m}$]{
    \includegraphics[width=1.5in]{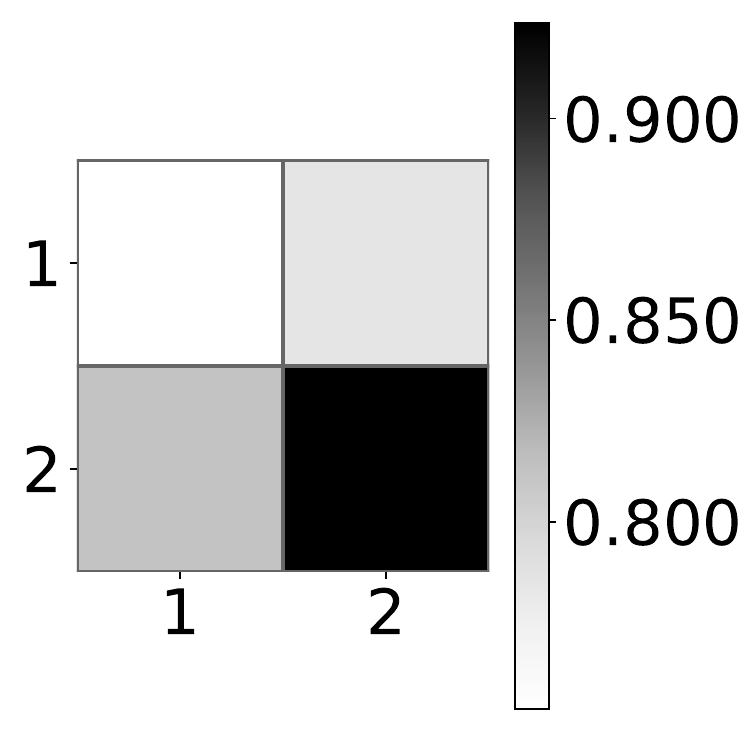}
    \label{fig:enronChpM}
    }%
    \subfigure[$\hat{\mu}/(1-\hat{m})$]{
    \includegraphics[width=1.5in]{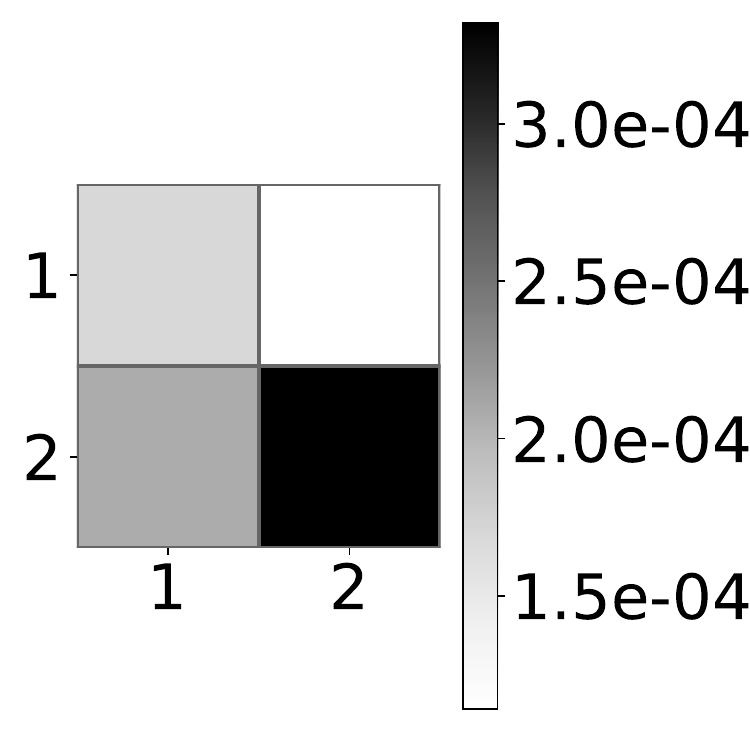}
    \label{fig:enronChpMuOver1M}
    }
    \caption{
    Estimated CHIP parameters on Enron data, where axis labels of each heatmap denote block index. Each tile corresponds to a block pair where $(a,b)$ denotes row $a$ and column $b$.}
\label{fig:enronChpFit}
\end{figure}

\begin{table}[tp]
    \centering
        \caption{Estimated $\hat{m}_{ab} \pm 95\%$ confidence interval from CHIP on the Enron dataset with $k=2$. All values of $\hat{m}_{ab}$ are statistically significant at the 5\% level for the test $m_{ab} > 0$. The high values of $\hat{m}_{ab}$ indicate that interactions in all block pairs are quite bursty.}
        \label{tab:enronMConfInterval}
    \begin{tabular}{r|cc}
        \toprule 
        \multicolumn{1}{c}{Block Pair} & 1    & 2    \\ \cline{2-3}
        \multicolumn{1}{r|}{1}     & 0.7536 $\pm$ 0.0440 & 0.7855 $\pm$ 0.0572 \\ \cline{2-3}
        \multicolumn{1}{r|}{2}     & 0.8126 $\pm$ 0.0424 & 0.9237 $\pm$ 0.0362 \\
        \bottomrule 
    \end{tabular}
\end{table}

\begin{table}[tp]
    \centering
    \caption{Pairwise difference for unique pairs of diagonal vs. off-diagonal $\hat{\mu}_{a, a} - \hat{\mu}_{a, b} \pm 95\%$ confidence interval of the CHIP model fitted to the Enron dataset with $k=2$. None of the differences are statistically significant at the 5\% level for the test $\hat{\mu}_{a, a} - \hat{\mu}_{a, b} \neq 0$, suggesting that the community structure is not evident from differences in the baseline rates $\mu$.}
    \label{tab:enronMuPairwiseDiff}
    \begin{tabular}{c|c}
        \multicolumn{2}{c}{Pairwise Differences in $\hat{\mu}$} \\
        \toprule
        $\hat{\mu}_{1, 1} - \hat{\mu}_{1, 2}$ & $1.724\times10^{-5} \pm 2.970\times10^{-5}$ \\ \hline
        $\hat{\mu}_{1, 1} - \hat{\mu}_{2, 1}$ & $2.929\times10^{-6} \pm 3.455\times10^{-5}$ \\ \hline
        $\hat{\mu}_{2, 2} - \hat{\mu}_{1, 2}$ & $8.650\times10^{-7} \pm 4.105\times10^{-5}$ \\ \hline
        $\hat{\mu}_{2, 2} - \hat{\mu}_{2, 1}$ & $-1.345\times10^{-5} \pm 4.468\times10^{-5}$ \\ \bottomrule
    \end{tabular}
\end{table}

Figure \ref{fig:enronChpMu} shows the estimated baseline intensity of each block pair. This can be thought of as the rate at which email conversations get started. 
We observe that $\hat{\mu}_{11}$ is much larger than $\hat{\mu}_{22}$; however block pair $(1, 1)$ only accounts for 965 emails as opposed to 1,425 for block pair $(2, 2)$. 
Thus, the community structure is not evident only from the differences in the baseline rates $\mu$.

It is only after we consider how bursty interactions are in each block pair, as shown in Figure \ref{fig:enronChpM}, that we can explain the dynamics of this network. 
In particular, $\hat{m}_{22}$ is much higher than $\hat{m}_{11}$.
In other words, once an email conversation is started in block-pair $(2, 2)$ we can expect more emails to follow, as opposed to more frequent conversations starting in $(1, 1)$, but with less follow-ups.
Hence, the combination of $\mu$ and $m$ allows us to observe the community structure, with more edges within block pairs than between, as shown by the values of $\hat{\mu}/(1-\hat{m})$ in Figure \ref{fig:enronChpMuOver1M}.

Table \ref{tab:enronMConfInterval} shows the numerical values for $\hat{m}$ along with their 95\% confidence intervals obtained using \eqref{eq:m_conf_int}, indicating that all block pairs exhibit highly bursty behavior.
As previously mentioned, the baseline rates $\hat{\mu}$ are not by themselves indicative of the community structure due to the burstiness of events in all of the blocks. 
Indeed, when we examine the 95\% confidence intervals for pairwise differences between the $\mu$ values for different block pairs using \eqref{eq:mu_diff_conf_int} shown in Table \ref{tab:enronMuPairwiseDiff}, all of the confidence intervals include 0.

\subsubsection{Exploratory Analysis of Facebook Wall Post Network} \label{sec:fbExploratoryAnalysis}

\begin{figure}[t]
    \centering
    \subfigure[Singular values]{
    \includegraphics[height=2in]{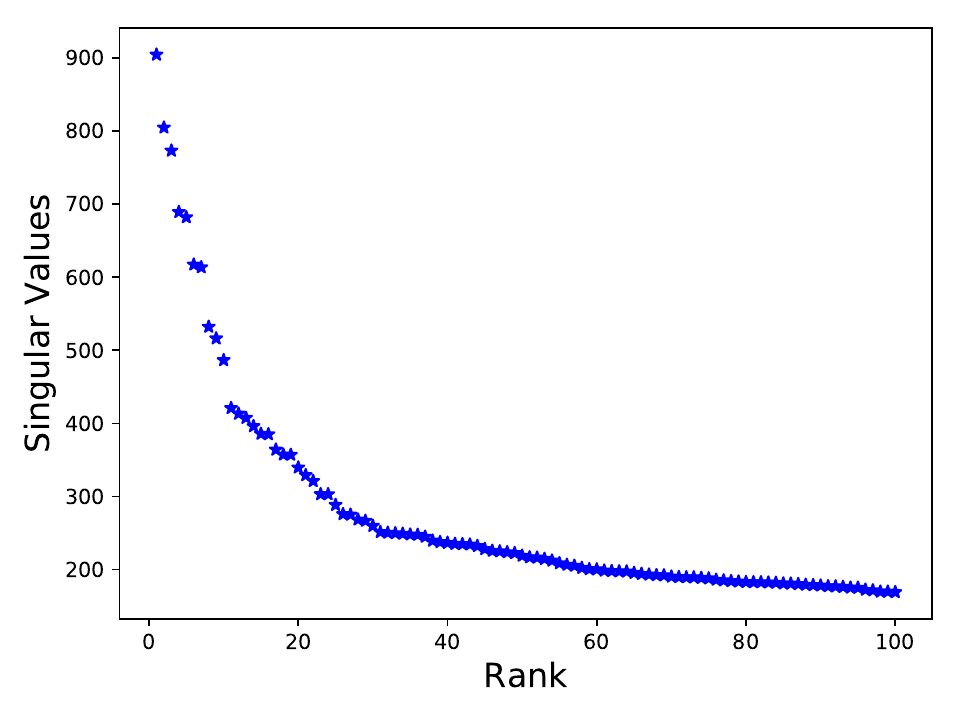}
    \label{fig:fbChpSingularValues}
    }
    \subfigure[Block sizes]{
    \includegraphics[height=2.2in]{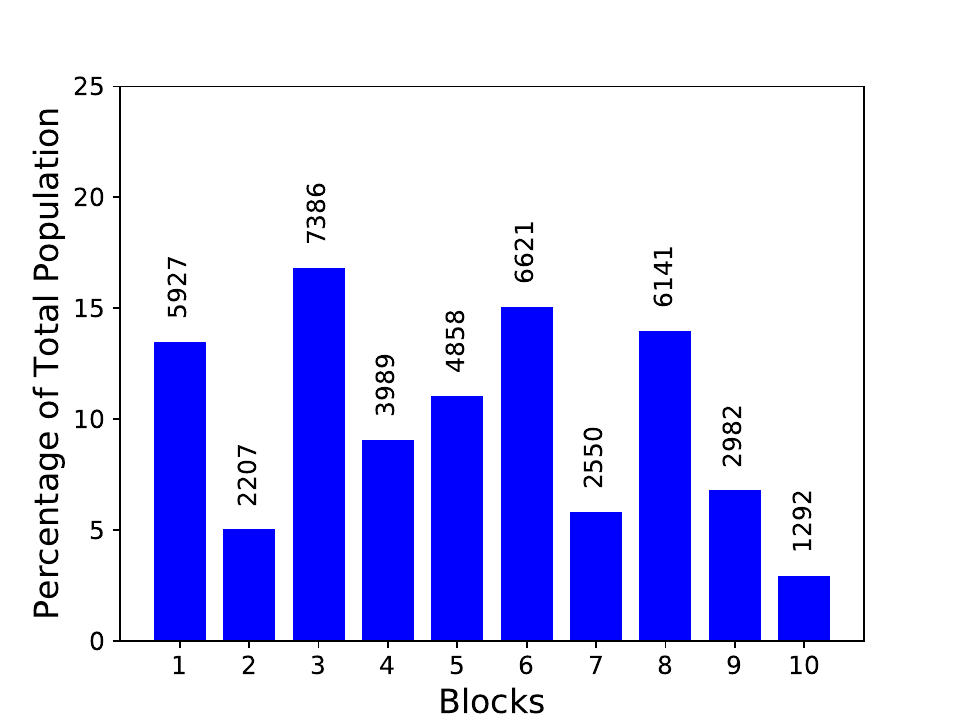}
    \label{fig:fbChpBlockSize}
    }
    \caption[Results of spectral clustering on the weighted adjacency matrix of the largest connected component of the Facebook Wall Posts dataset]
    {Results of spectral clustering on the weighted adjacency matrix of the largest connected component of the Facebook Wall Posts dataset. \subref{fig:fbChpSingularValues} 100 largest singular values. There is a large gap between the 10\textsuperscript{th} and the 11\textsuperscript{th} largest singular values that leads us to select $k=10$ blocks. \subref{fig:fbChpBlockSize} Size of each formed block. Numbers on top of each bar indicate the actual number of nodes in that block.}
\label{fig:fbChpSc}
\end{figure}

\begin{figure}[t!]
    \centering
    \includegraphics[width=3.5in]{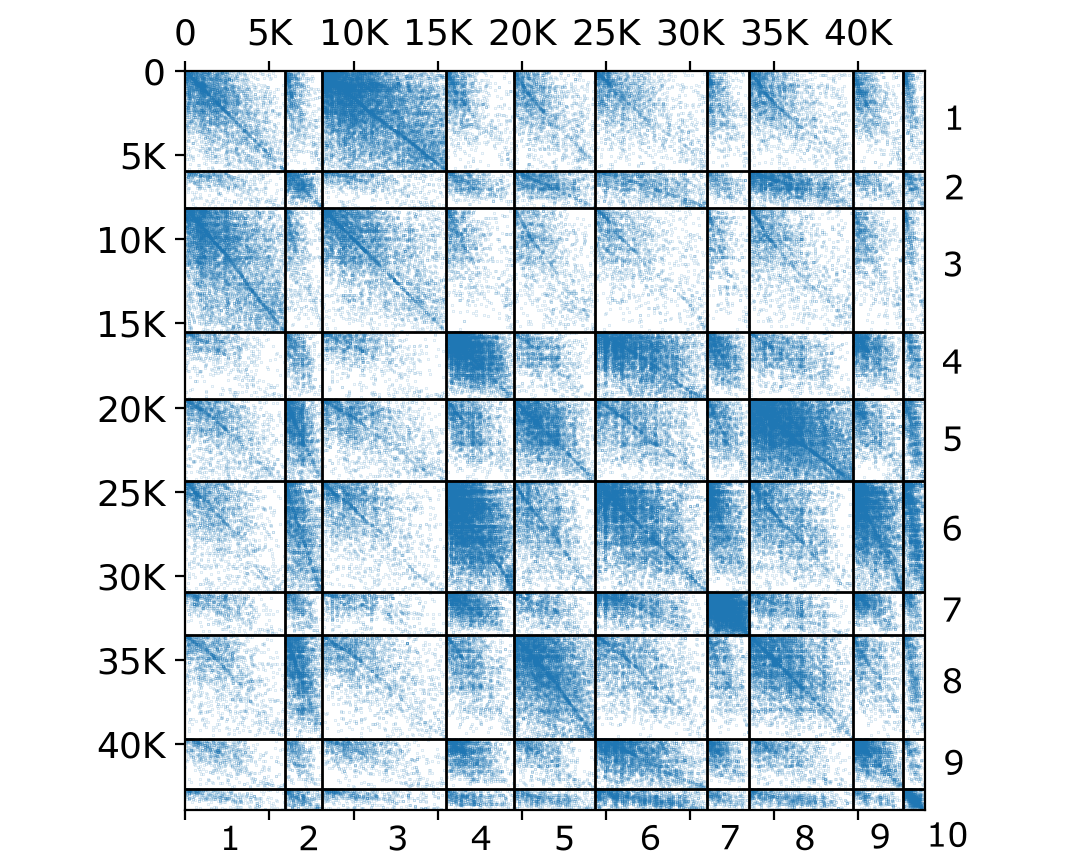}
    \caption{Adjacency matrix for Facebook wall post network with rows and columns rearranged to show block structure.}
    \label{fig:fbChpAdjMatSupp}
\end{figure}

Fitting the CHIP model to the largest connected component of the network (excluding self loops) consisting of 43,953 nodes and 852,833 edges required only 141.4 s.
Considering the gap between the 10\textsuperscript{th} and the 11\textsuperscript{th} largest singular values of the weighted adjacency matrix of the network as shown in Figure \ref{fig:fbChpSingularValues}, we choose a model with $k=10$ blocks, resulting in the block sizes depicted in Figure \ref{fig:fbChpBlockSize}.

\begin{figure}[tp]
    \centering
    \subfigure[$\mu$: base intensity]{
    \includegraphics[width=2.3in]{fb_mu}
    \label{fig:fbChpMuSupp}
    }
    \subfigure[$m$: $\alpha$ to $\beta$ ratio]{
    \includegraphics[width=2.3in]{fb_m}
    \label{fig:fbChpMSupp}
    }
    \quad
    \subfigure[$\alpha$: intensity jump size]{
    \includegraphics[width=2.3in]{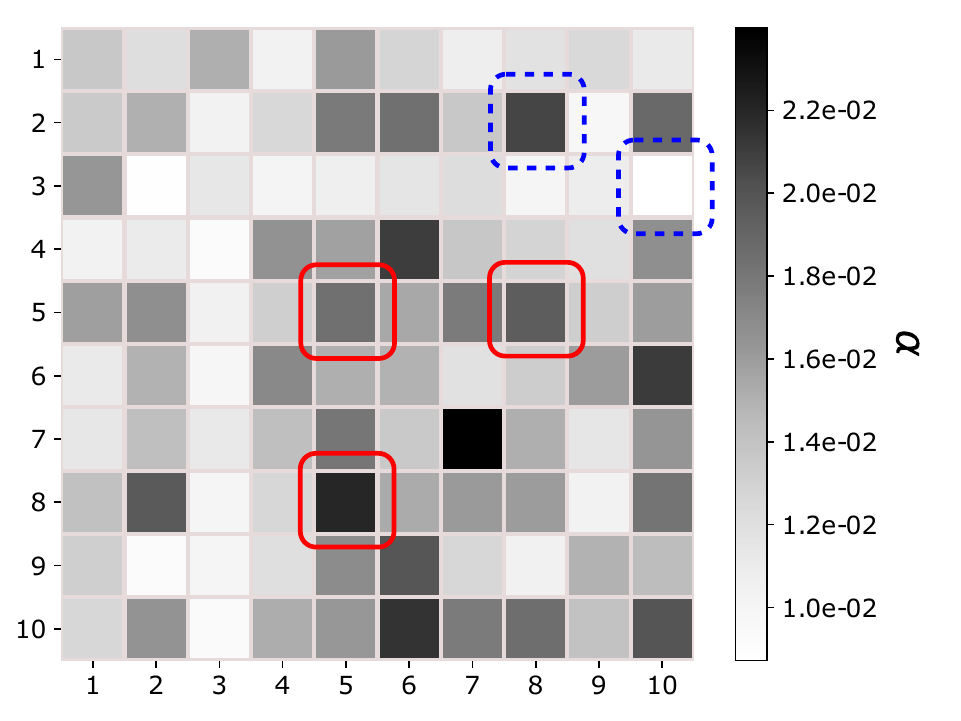}
    \label{fig:fbChpAlpha}
    }
    \subfigure[$\beta$: intensity decay rate]{
    \includegraphics[width=2.3in]{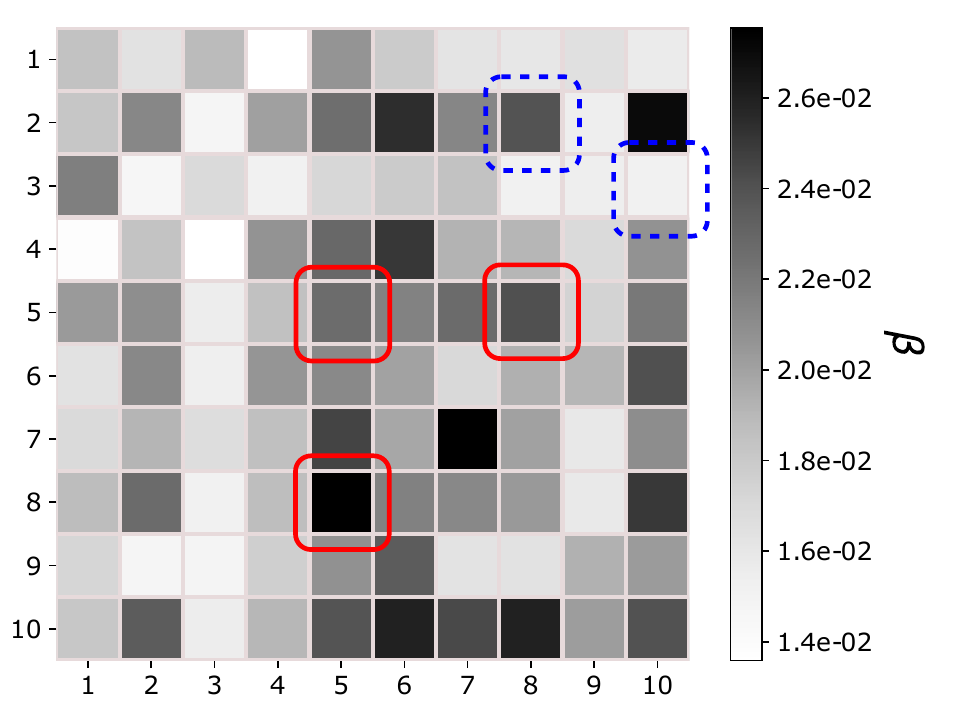}
    \label{fig:fbChpBeta}
    }
    \quad
    \subfigure[Total number of events]{
    \includegraphics[width=2.3in]{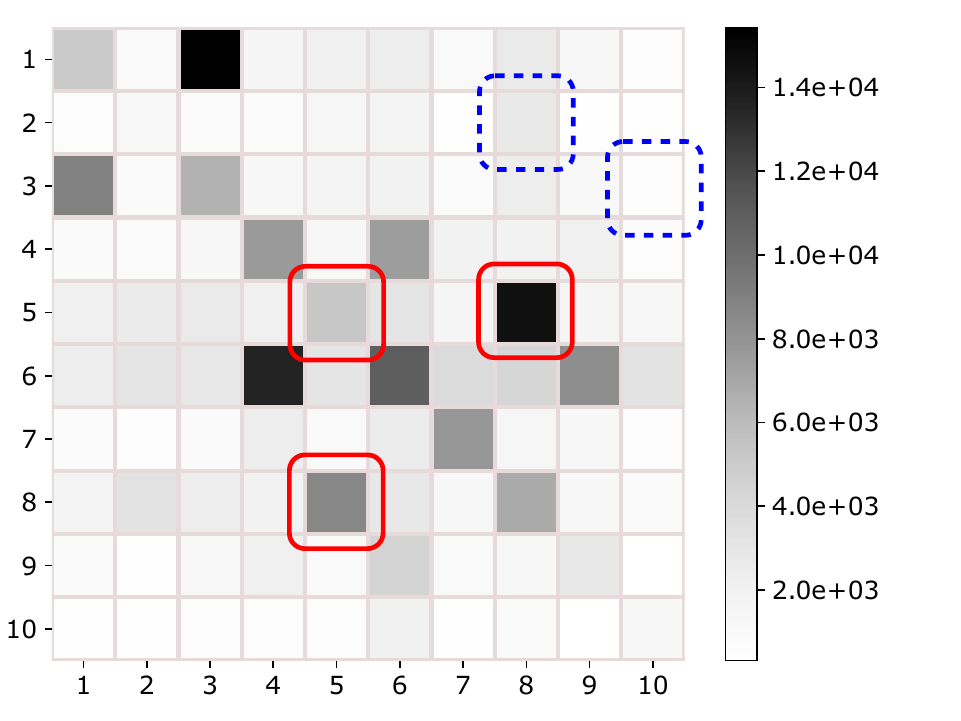}
    \label{fig:fbChpTotalNumEvents}
    }
    \subfigure[Mean number of events per node pair]{
    \includegraphics[width=2.3in]{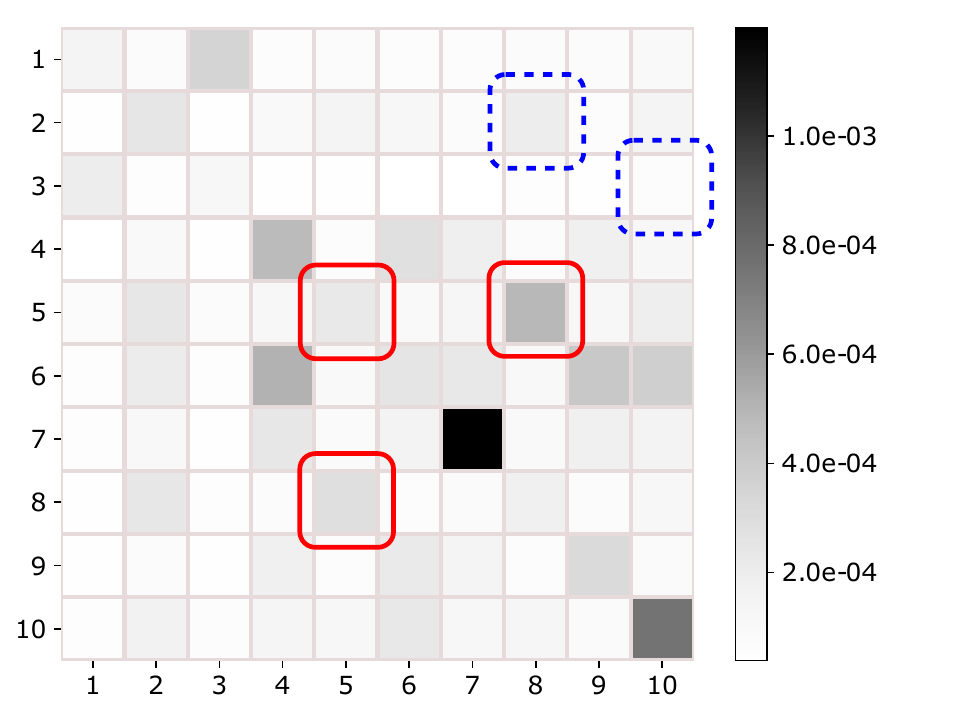}
    \label{fig:fbChpNumEventsNodePair}
    }
    \caption{Inferred CHIP parameters on the largest connected component of the Facebook Wall Posts dataset with $k=10$. Axis labels denote block numbers. Each tile corresponds to a block pair where $(a,b)$ denotes row $a$ and column $b$. Boxed block pairs in the heatmap are discussed in the body text.}
\label{fig:fbChpFitSupp}
\end{figure}

Figure \ref{fig:fbChpFitSupp} shows heatmaps of the fitted CHIP parameters. 
Although diagonal block pairs have a higher base intensity on average, indicating an underlying assortative community structure, there are some off-diagonal block pairs with a high $\mu$ such as $(5, 8)$ and $(8, 5)$, as shown in the red boxes in Figure \ref{fig:fbChpFitSupp}.
This illustrates that the CHIP model does not discourage inter-block events. These patterns often occur in social networks, for instance, if there are communities with opposite views on a particular subject.

While the structure of $\mu$ reveals insights on the baseline rates of events between block pairs, the structure of $\alpha$ (Figure \ref{fig:fbChpAlpha}) and $\beta$ (Figure \ref{fig:fbChpBeta}) reveal insights on the burstiness of events between block pairs. 
Note that the structure of the $\alpha$ to $\beta$ ratio $m$ (Figure \ref{fig:fbChpM}) affects the asymptotic mean number of events in \eqref{eq:asy_mean}. 
For some block pairs, such as $(3,10)$, there are very low values of $\alpha$ and $\beta$ indicating the events are closely approximated by a homogeneous Poisson process.
There are some block pairs, such as $(2,8)$ that have a low baseline rate of events but are extremely bursty, which relatively increases the mean number of events per node pair. 
Both of these block pairs are shown in the blue dashed boxes in Figure \ref{fig:fbChpFitSupp}.
The different levels of burstiness of block pairs cannot be seen from aggregate statistics such as the total number of events (Figure \ref{fig:fbChpTotalNumEvents}) or even the mean number of events per node pair (Figure \ref{fig:fbChpNumEventsNodePair}).

Unlike the findings of \cite{junuthula2017block}, who studied only a subset of the network containing $3,582$ nodes using $k=2$ blocks, we find that $\alpha$ is not necessarily higher for diagonal blocks as shown in Figure \ref{fig:fbChpAlpha}.
Additionally, even though we do not explicitly model reciprocity between node pairs in our CHIP model, we can nevertheless empirically observe certain reciprocities through the patterns of the estimated $\alpha$ and $\beta$ parameters.
We note that the high reciprocity present in social networks is captured by CHIP through the symmetry in all Hawkes process parameters about block pairs. This can be observed in block pairs $(8,5)$ and $(5,8)$. In the context of this dataset, a symmetric $\alpha$ and $\beta$ corresponds to the notion that wall posts posted by the people in block $5$ on the wall of the people in block $8$ will urge people in block $8$ to respond, which in turn promotes more wall posts by people in block $5$. 

Lastly, it is worth noting that fitting the CHIP model to this data set using the unweighted adjacency matrix resulted in a per event log-likelihood of $-10.04$ compared to $-9.61$ for the weighted adjacency matrix on the test data set when using a 80\%/20\% train and test split on the events. Thus, this was another reason to use the weighted adjacency matrix besides its aforementioned advantages in previous sections. We note that running spectral clustering on the unweighted adjacency matrix, compared to the weighted adjacency matrix, seemed to detect communities with larger number of intra-block events, while inter-block events were a lot less common.

\bibliography{references}
\bibliographystyle{icml2020}

\makeatletter\@input{xxms.tex}\makeatother